\documentclass[%%
a4paper,%           % letterpaper, a4paper, a5paper               %
10pt,%               % 10pt,11pt, 12pt                             %
superscriptaddress,% % authors with affiliations via superscripts  %
amsmath,%            % add AMS-Latex features                   	 %
amssymb,%            % add extra AMS symbols, including amsfonts   %
aps,%				 %												 %
prd,%       	     % prl, pra, prb, prc, prd, pre, prstab        %
showpacs,% 			 % make PACS codes appear                      %
]{revtex4}%         %                                           	 %

\usepackage{tensor}      % manipulate tensors				       %
\usepackage{graphicx}    % include figures  					   %
\usepackage[colorlinks=true,%        color link				   %
citecolor=blue,%		  % cite color							   %
linkcolor=blue,%		  % link color							   %
urlcolor=blue%			  % url color							   %
]{hyperref}				  % create hyperlinks		 			   %
\usepackage{bm}		      % \bm{<text>} Bold math symbols   	   %
\usepackage{dcolumn}	  % Align table columns on decimal point  %
\usepackage{array}	  	  % dcolumn depends on array			   %
\newcommand*{\be}{\begin{equation}}
\newcommand*{\ee}{\end{equation}}
\newcommand*{\bea}{\begin{eqnarray}}
\newcommand*{\eea}{\end{eqnarray}}
\def\disp{\displaystyle}

\begin{document}

%%%%%%%%%%%%%%%%%%%%%%%%%%%%%%%%%%%%%%%%%%%%%%%%%%%%%%%%%%%%%%%%%%%%%%
%%%%%%%%%%%%%%%%%%%%%%%%%%% the front matter %%%%%%%%%%%%%%%%%%%%%%%%%
%%%%%%%%%%%%%%%%%%%%%%%%%%%%%%%%%%%%%%%%%%%%%%%%%%%%%%%%%%%%%%%%%%%%%%
\begin{titlepage}

\begin{flushright}
arXiv:1503.05281
\end{flushright}

\title{\Large \bf \texorpdfstring{$f(T)$}{f(T)} non-linear massive
gravity and the cosmic acceleration}

%%%%%%%%%%%%%%%%%%%%%%%%%%%%%%%%%%%%%%%%%%%%%%%%%%%%%%%%%%%%%%%%%%%%%%
%%%%%%%%%%%%%%%%%%%%%%%%%%%%%%% authors %%%%%%%%%%%%%%%%%%%%%%%%%%%%%%
\author{You~Wu\,}
\email[\,Email address:\ ]{5u@ruc.edu.cn}
\affiliation{Department of Physics,
Renmin University of China,
Beijing 100872, China}

\author{Zu-Cheng~Chen\,}
\email[\,Email address:\ ]{bingining@gmail.com}
\affiliation{School of Physics,
Beijing Institute of Technology,
Beijing 100081, China}

\author{Jiaxin~Wang\,}
\email[\,Email address:\ ]{jxw@mail.nankai.edu.cn}
\affiliation{Department of Physics,
Nankai University,
Tianjin 300071, China}
\affiliation{State Key Laboratory of Theoretical Physics,
Institute of Theoretical Physics,
Chinese Academy of Science,
Beijing 100190, China}

\author{Hao~Wei\,}
\thanks{\,Corresponding author}
\email[\,Email address:\ ]{haowei@bit.edu.cn}
\affiliation{School of Physics,
Beijing Institute of Technology,
Beijing 100081, China}

%%%%%%%%%%%%%%%%%%%%%%%%%%%%%%%%%%%%%%%%%%%%%%%%%%%%%%%%%%%%%%%%%%%%%%
%%%%%%%%%%%%%%%%%%%%%%%%%%%%% abstract %%%%%%%%%%%%%%%%%%%%%%%%%%%%%%%
\begin{abstract}\vspace{1cm}
\centerline{\bf ABSTRACT}\vspace{2mm}
Inspired by the $f(R)$ non-linear massive gravity, we propose
a new kind of modified gravity model, namely $f(T)$ non-linear massive
gravity, by adding the dRGT mass term reformulated in the
vierbein formalism, to the $f(T)$ theory.
We then investigate the cosmological evolution of $f(T)$ massive
gravity, and constrain it by using the latest observational data.
We find that it slightly favors a crossing of the phantom
divide line from the quintessence-like phase ($w_{de} > -1$) to
the phantom-like one ($w_{de} < -1$) as redshift decreases.
\end{abstract}

% see http://www.aip.org/pacs

\maketitle

\end{titlepage}

%%%%%%%%%%%%%%%%%%%%%%%%%%%%%%%%%%%%%%%%%%%%%%%%%%%%%%%%%%%%%%%%%%%%%%
%%%%%%%%%%%%%%%%%%%%%%%%%%%%% section 1 %%%%%%%%%%%%%%%%%%%%%%%%%%%%%%
%%%%%%%%%%%%%%%%%%%%%%%%%%%%%%%%%%%%%%%%%%%%%%%%%%%%%%%%%%%%%%%%%%%%%%
\section{\label{Intro}Introduction}
Einstein's theory of general relativity (GR)
has achieved great success in explaining various gravitational
phenomena since its birth, and till today it can pass the Solar
System's tests to a very high precision.
However, on scales larger than the Solar System, there are some
astrophysical observations having not yet been fully understood or
explained in the context of GR.
Maybe the most urgent among them are the presence of
an invisible (dark) matter component in our universe,
first discovered by Zwicky in the $1930$s
\cite{Zwicky:1933gu,1937ApJ},
as well as the late-time acceleration
of our universe confirmed by various cosmological observations
\cite{Riess:1998cb,Perlmutter:1998np,Spergel:2003cb,%%
Tegmark:2003ud,Allen:2004cd,Riess:2004nr,Spergel:2006hy,%%
Ade:2013ktc,Ade:2013zuv}.
Although by extending the standard model of particle physics,
there exist various promising candidates for dark matter
(see e.g. reviews \cite{Steffen:2008qp,Bergstrom:2009ib,Feng:2010gw}),
so far there has been no natural and satisfactory explanation
for the cosmic acceleration by introducing some kinds of
mysterious cosmological fluids with negative pressure
called dark energy (DE), since they all suffer from the fine-tuning
\cite{Weinberg:1988cp} and/or the cosmic coincidence problems
\cite{fitch1997critical,Peebles:2002gy,Copeland:2006wr}.
Unless all of these problems are eventually solved on the
particle physics side, the severity of the phenomenological problems
challenges GR as the ultimate and complete theory of gravity.
Furthermore, difficulties arise when trying to search for a
quantum theory of gravity, since GR is not renormalizable
\cite{tHooft:1974bx,Goroff:1985th}
and thus cannot make meaningful physical predictions.
It is, therefore, reasonable to investigate the possibility of
modifying Einstein's gravity theory.

Among the modified gravity theories, $f(R)$ theory
(see e.g. \cite{Sotiriou:2008rp,DeFelice:2010aj,Nojiri:2006ri,
Nojiri:2010wj} for recent reviews) serves as one
of the most straightforward and popular generalizations of GR by
extending the Ricci scalar $R$ in the Einstein-Hilbert action
to a general function $f(R)$.
It has been shown that $f(R)$ gravity can explain the present
cosmic acceleration without the need of DE
\cite{Amendola:2007nt,Appleby:2007vb,Starobinsky:2007hu,%
Linder:2009jz,Cognola:2007zu,Deruelle:2008fs,Tsujikawa:2007xu,%%
Hu:2007nk,Li:2007xn,Amendola:2006we,Nojiri:2007cq,Nojiri:2007as}.
But in general, the resulting field equations
of motion are $4$th order in $f(R)$ gravity.
This feature makes $f(R)$ theory quite hard to analyse.
In comparison with $f(R)$ gravity, there is another kind of
modified gravity, namely $f(T)$ theory,
which has been extensively explored
in the literature (see e.g. \cite{Bengochea:2008gz,Linder:2010py,%
Wei:2011jw,Wei:2011mq,Wei:2011aa,Yang:2010ji,Li:2010cg,%
Li:2011wu,Wu:2010av,Nashed:2014lva,Li:2011rn,Nesseris:2013jea}).
Here $T$ is the torsion scalar constructed
from the curvatureless Weitzenb\"ock connection
\cite{weitzenbock1923invariantentheorie}.
It is well known that $f(T)$ theory is based on the teleparallel
space-time, in which vierbein field is used as the basic dynamical
variable instead of metric in GR.
One advantage of $f(T)$ theory in contrast to $f(R)$ theory is
that, the field equations arising from $f(T)$ theory are $2$nd order,
and thus simpler than $f(R)$ case.

On the other hand, as a classical field theory, GR propagates a
non-linear self-interacting massless spin-$2$ field.
Therefore, another way to modify GR is to give a tiny mass to the
graviton.
This class of theories
is known as massive gravity in the literature.
Actually, massive gravity has a long and elusive history.
The first attempt to give a mass to graviton dates back to the year
of 1939 when Fierz and Pauli proposed a unique ghost-free mass
term to gravity in the linear perturbation level \cite{Fierz:1939ix}.
The linearised massive gravity propagates $5$ degrees of freedom
(DoF).
And then, in 1970s, van Dam-Veltman-Zakharov (vDVZ) discontinuity was
found. It was shown that, when the mass of graviton $m \rightarrow 0$,
Fierz-Pauli theory would not converge to massless case
\cite{vanDam:1970vg,Zakharov:1970cc},
thus giving different predictions than those GR.
Almost at the same time, Vainshtein \cite{Vainshtein:1972sx} showed
that the linear approximation loses its validity around massive
sources below the Vainshtein radius and non-linear effects
should be considered.
The vDVZ discontinuity can then be avoided by this well-known
Vainshtein screening mechanism.
However, it was claimed that any non-linear extensions of
Fierz-Pauli theory will generally have $6$ DoF, with the extra DoF
being a Boulware-Deser (BD) ghost \cite{Boulware:1973my}.
The construction of a well-defined
non-linear theory of massive gravity without the BD ghost
has been a challenging problem over decades.
Recently, it was shown that BD ghost may be removed order-by-order
in a careful construction of mass term up to a certain decoupling
limit~\cite{deRham:2010ik}.
The full non-linear massive gravity was resummed by
de Rham, Gabadadze and Tolley (dRGT) in \cite{deRham:2010kj},
and the absence of BD ghost was further confirmed by using Hamiltonian
constraint analysis in \cite{Hassan:2011ea,Hassan:2011hr}.
Although dRGT theory exists accelerating solutions
\cite{Motohashi:2012jd,Gumrukcuoglu:2012aa,Kobayashi:2012fz,
D'Amico:2011jj},
it suffers from some
instabilities at the cosmological perturbation level
\cite{DeFelice:2012mx,DeFelice:2013bxa,DeFelice:2013awa}.
Since then dRGT theory has been extended in various approaches,
e.g. the mass varying massive gravity \cite{Huang:2012pe,Wu:2013ii}
and the quasi-dilaton massive
gravity \cite{D'Amico:2012zv,D'Amico:2013kya,DeFelice:2013dua}.

Recently, a new extension of massive gravity, namely $f(R)$
non-linear massive gravity, has been proposed
\cite{Cai:2013lqa,Cai:2014upa}.
It is shown that this model not only shares no linear
instabilities around a cosmological background,
but can also describe the inflation and the recently cosmic
acceleration in a unified framework \cite{Cai:2013lqa,
Cai:2014upa,Kluson:2013yaa}. Inspired by the $f(R)$ non-linear
massive gravity, we propose a new kind of modified gravity
theory by adding the non-linear dRGT mass term formulated in
the vierbein formalism, to the $f(T)$ theory.
We might call this theory $f(T)$ non-linear massive gravity,
or simply $f(T)$ massive gravity.
One merit of our model compared with the $f(R)$
massive gravity is that, the field equations
arising from $f(T)$ massive gravity are $2$nd order,
and thus much easier to analyse.
Moreover, $f(R)$ massive gravity, as well as dRGT theory,
is generally constructed in the metric formulation.
However, in the metric
formalism, the mass term is difficult to analyse, since it
contains the square root of the metric.
Another merit of our theory is that, in teleparallel space-time,
we use vierbein instead of metric as the basic dynamical variable.
Since vierbein can be viewed as the square root of the metric
in some sense, the vierbein formalism in our theory can
greatly simplify the calculation.

After presenting the action of our model,
we then investigate the cosmological evolution of $f(T)$
massive gravity.
The rest of this paper is organized as follows.
In Sec.~\ref{ftGravity}, we briefly review the $f(T)$ gravity.
In Sec.~\ref{ftMassiveGravity}, we add the ghost-free
dRGT mass term
reformulated in the vierbein formalism to the $f(T)$ sector,
thus giving the action of $f(T)$ massive gravity.
The evolutionary equations for the flat Friedmann-Robertson-Walker
(FRW) cosmology are given as well.
In Sec.~\ref{observations}, we focus on two concrete examples,
namely the power-law and exponential types
of $f(T)$ massive gravity and investigate their
cosmological evolution.
We constrain the model parameters by the recent cosmological data
and study the phantom crossing behavior.
In Sec.~\ref{Conclusions}, we give some concluding remarks.

%%%%%%%%%%%%%%%%%%%%%%%%%%%%%%%%%%%%%%%%%%%%%%%%%%%%%%%%%%%%%%%%%%%%%%
%%%%%%%%%%%%%%%%%%%%%%%%%%%%% section 2 %%%%%%%%%%%%%%%%%%%%%%%%%%%%%%
%%%%%%%%%%%%%%%%%%%%%%%%%%%%%%%%%%%%%%%%%%%%%%%%%%%%%%%%%%%%%%%%%%%%%%
\section{\label{ftGravity}\texorpdfstring{$f(T)$}{f(T)} gravity}

$f(T)$ gravity, which is a generalization of the teleparallel
gravity originally proposed by Einstein
\cite{einstein1928riemann,einstein1928neue,einstein1930auf,
Unzicker:2005in,Hayashi:1979qx},
has received great interest in the literature recently.
Following \cite{Ferraro:2006jd,Ferraro:2008ey,Bengochea:2008gz},
we now briefly review the $f(T)$ theory.
$f(T)$ gravity is built upon teleparallel space-time.
In teleparallel space-time, the basic dynamical
quantity is a vierbein field
$e_a = \tensor{e}{_a^\mu} \partial_{\mu}$, with Latin indices
$a, b, \cdots = 0, \cdots, 3$,
and $i, j, \cdots = 1, \cdots, 3$, Greek indices
$\mu, \nu, \cdots = 0, \cdots ,3$, and $\partial_{\mu}$
coordinate bases. We also note that the Einstein summation
notation for the indices is used throughout this paper.
The vierbein is an orthonormal basis for
the tangent space at each point $x^{\mu}$ of the manifold,
namely $e_a \cdot e_b = \eta_{a b}$, with
$\eta_{a b} = \mathrm{diag}\,(-1,\, 1,\, 1,\, 1)$.
The metric tensor can then be expressed in the dual vierbein
$\tensor{e}{^a_\mu}$ as
\be\label{gee}
	g_{\mu \nu}(x)
	= \eta_{a b}\, \tensor{e}{^a_\mu}(x)\, \tensor{e}{^b_\nu}(x) .
\ee
Rather than using the torsionless Levi-Civita connection
in general relativity (GR), teleparallel space-time uses
the curvatureless Weitzenb\"{o}ck connection
$\tensor{\Gamma}{^\lambda_\mu_\nu}$,
which is defined by
\be
	\tensor{\Gamma}{^\lambda_\mu_\nu}
	\equiv \tensor{e}{_a^\lambda} \partial_{\mu} \tensor{e}{^a_\nu} .
\ee
Note that the lower indices $\mu$ and $\nu$ are not symmetric
in general, thus the torsion tensor is non-zero in the
teleparallel space-time. The Weitzenb\"{o}ck torsion tensor
is defined by
\be
	\tensor{T}{^\lambda_\mu_\nu}
	\equiv \tensor{\Gamma}{^\lambda_\nu_\mu}
		- \tensor{\Gamma}{^\lambda_\mu_\nu}
	= \tensor{e}{_a^\lambda} \left(
			\partial_{\nu} \tensor{e}{^a_\mu}
			- \partial_{\mu} \tensor{e}{^a_\nu}
			\right) .
\ee
In teleparallel gravity, the gravitational action is given by
the torsion scalar instead of the the Ricci scalar in GR.
The torsion scalar is basically the square of the
Weitzenb\"{o}ck torsion tensor, and is defined by
\be
	T
	\equiv \tensor{S}{^\rho_\mu_\nu} \tensor{T}{_\rho^\mu^\nu}
	= \frac{1}{4} \tensor{T}{^\rho_\mu_\nu} \tensor{T}{_\rho^\mu^\nu}
		+ \frac{1}{2} \tensor{T}{^\rho_\mu_\nu}
			\tensor{T}{^\nu^\mu_\rho}
		- \tensor{T}{^\rho_\mu_\rho} \tensor{T}{^\nu^\mu_\nu} ,
\ee
with the tensor $\tensor{S}{^\rho_\mu_\nu}$ given by
\be
	\tensor{S}{^\rho_\mu_\nu}
	\equiv \frac{1}{4} \left(
			\tensor{T}{^\rho_\mu_\nu}
			- \tensor{T}{_\mu_\nu^\rho}
			+ \tensor{T}{_\nu_\mu^\rho}
			\right)
		+ \frac{1}{2} \delta^\rho_\mu \tensor{T}{^\sigma_\nu_\sigma}
		- \frac{1}{2} \delta^\rho_\nu \tensor{T}{^\sigma_\mu_\sigma} .
\ee

In $f(T)$ gravity, the gravitational field is driven by a
Lagrangian density $T + f(T)$, with $f(T)$ a function of $T$,
and the action reads
\be
	S = -M_p^2 \int e \left[ T + f(T) \right] d^4x
		+ S_m\left[ \tensor{e}{_a^\mu}, \chi_m \right] ,
\ee
where $e = \mathrm{det}\,(\tensor{e}{^a_\mu}) = \sqrt{-g}$,
$M_p^2 = (16 \pi G)^{-1}$, with $g$ the determinant of the metric
$g_{\mu\nu}$ and $G$ the Newtonian constant.
Note that we have used the units in which the speed of light
$c = 1$, and the reduced Planck constant $\hbar = 1$.
Here, $S_m\left[ \tensor{e}{_a^\mu}, \chi_m \right]$ is the
matter part of the action, and $\chi_m$ denotes all matter
fields collectively.
If we set $f(T) = 0$, then the action gives an
equivalent description of the space-time as GR.

%%%%%%%%%%%%%%%%%%%%%%%%%%%%%%%%%%%%%%%%%%%%%%%%%%%%%%%%%%%%%%%%%%%%%%
%%%%%%%%%%%%%%%%%%%%%%%%%%%%% section 3 %%%%%%%%%%%%%%%%%%%%%%%%%%%%%%
%%%%%%%%%%%%%%%%%%%%%%%%%%%%%%%%%%%%%%%%%%%%%%%%%%%%%%%%%%%%%%%%%%%%%%
\section{\label{ftMassiveGravity}
\texorpdfstring{$f(T)$}{f(T)} non-linear massive gravity}

In this section, we will add the dRGT mass term reformulated in the
vierbein form to the $f(T)$ sector, thus giving the action of $f(T)$
massive gravity. We then derive the evolutionary equations for the flat
FRW cosmology.

\vspace{-3mm} % used here just for a better typesetting

%%%%%%%%%%%%%%%%%%%%%%%%%%%%%%%%%%%%%%%%%%%%%%%%%%%%%%%%%%%%%%%%%%%%%%
%%%%%%%%%%%%%%%%%%%%%%%%%%%%%%%%%%%%%%%%%%%%%%%%%%%%%%%%%%%%%%%%%%%%%%
\subsection{The action}

Since teleparallel gravity equals to GR at the field equations
level, here we give an equivalent description of dRGT massive theory
by adding the mass term to teleparallel gravity.
The mass term firstly proposed in dRGT theory,
can be greatly simplified when transforming to the vierbein formalism.
We refer to \cite{Hinterbichler:2012cn} for a detailed derivation.
By adding the mass term to teleparallel gravity,
one will get the teleparallel version of ghost-free dRGT
massive gravity as
\be\label{TMassiveGravity}
	S = -M_p^2 \int e\, T d^4x
		- M_p^2 m^2 \int \sum_{n=0}^{4} e\, \beta_n S_n
              \left(\mathbb{E}\right) d^4x
		+ S_m\left[ \tensor{e}{_a^\mu}, \chi_m \right] ,
\ee
where $\beta_n$ are free parameters,
$\mathbb{E}$ denotes the matrix of vierbein $\tensor{e}{_a^\mu}$
and $S_n(\mathbb{M})$ are the so-called elementary
symmetric polynomials \cite{Hinterbichler:2012cn}.
For an arbitrary $4 \times 4$ matrix $\mathbb{M}$,
the first few of $S_n(\mathbb{M})$ can be written in
the form \cite{Hinterbichler:2012cn}
\bea
	S_1(\mathbb{M})
		&=& [\mathbb{M}] ,\\[1mm]
	S_2(\mathbb{M})
		&=& \frac{1}{2!} \left( [\mathbb{M}]^2 - [\mathbb{M}^2]
              \right) ,\\[1mm]
	S_3(\mathbb{M})
		&=& \frac{1}{3!} \left(
			[\mathbb{M}]^3
			- 3[\mathbb{M}][\mathbb{M}^2]
			+ 2[\mathbb{M}^3]
			\right) ,
\eea
with $[\mathbb{M}]$ the trace of the matrix $\mathbb{M}$.
By generalizing the torsion scalar $T$ to a function $T + f(T)$
in the action~\eqref{TMassiveGravity}, we have the $f(T)$ non-linear
massive gravity as
\be\label{action}
	S = -M_p^2 \int e \left[ T + f(T) \right] d^4x
		- M_p^2 m^2 \int \sum_{n=1}^{3} e \beta_n S_n
              \left(\mathbb{E}\right) d^4x
		+ S_m\left[ \tensor{e}{_a^\mu}, \chi_m \right] ,
\ee
Note that we have absorbed the cosmological constant arising from the
mass term to the $f(T)$ sector. So, there are only three free
parameters
$\beta_1$, $\beta_2$, and $\beta_3$ in the mass term.
We also note that when $f(T) = 0$, the action~\eqref{action} is
just the teleparallel gravity plus a dRGT mass term
reformulated in the vierbein formalism, thus being equivalent to the
original dRGT theory at the field equations level.
So, if $f(T) = 0$, then the BD ghost will not show up in the
action~\eqref{action}.
However, if $f(T)$ is a non-trivial function of $T$, then additional
DoF will be introduced.
Therefore, we should perform Hamiltonian analysis to
confirm that the action~\eqref{action} is ghost-free.
In this paper we focus on the cosmological behavior of
$f(T)$ non-linear massive gravity, and the ghost problem
or other instabilities that our theory may suffer from,
will be considered in our future work.

%%%%%%%%%%%%%%%%%%%%%%%%%%%%%%%%%%%%%%%%%%%%%%%%%%%%%%%%%%%%%%%%%%%%%%
%%%%%%%%%%%%%%%%%%%%%%%%%%%%%%%%%%%%%%%%%%%%%%%%%%%%%%%%%%%%%%%%%%%%%%
\subsection{Equation of motion}

We now consider a spatially flat FRW space-time
\be
	ds^2 = -N^2(\tau)\, d^2\tau + a^2(\tau)\, \delta_{i j} dx^i dx^j ,
\ee
where $a(\tau)$ is the scale factor and $N(\tau)$ is the lapse
function.
This metric arises from the diagonal dual vierbein
\be
	\tensor{e}{^a_\mu}
	= \mathrm{diag}\left( N(\tau),\, a(\tau),\, a(\tau),\, a(\tau) \right)
\ee
through Eq.~\eqref{gee}.
Then we can easily get
$\tensor{e}{_a^\mu}
= \mathrm{diag}\left( N^{-1},\, a^{-1},\, a^{-1},\, a^{-1} \right)$,
$e = Na^3$, and hence
\bea
	S_1\left(\mathbb{E}\right) &=& \frac{3}{a} + \frac{1}{N} ,\\[0.5mm]
	S_2\left(\mathbb{E}\right) &=& \frac{3}{a^2} + \frac{3}{Na} ,\\[0.5mm]
	S_3\left(\mathbb{E}\right) &=& \frac{1}{a^3} + \frac{3}{Na^2} .
\eea
The torsion scalar is also obtained to be
$T = 6 a^{\prime\, 2} /(N^2 a^2) = 6 H^2$,
where a prime denotes $\frac{d}{d \tau}$.

After some algebra, action~\eqref{action} reduces to
\be\label{finalAction}
	S = M_p^2 \int \Big\{
			-N a^3 \left( f + T \right)
			- m^2 \big[
				\beta_1 \left( 3 N a^2 + a^3 \right)
				+ \beta_2 \left( 3 N a + 3 a^2 \right)
				+ \beta_3 \left( N + 3 a \right)
				\big]
			\Big\}\, d^4x
		+ S_m .
\ee
For further convenience, we define some variables here, namely
\bea
	x &=& a^{-1} ,\\
	y_1(x)&=& 3 \beta_1 x
			+ 3 \beta_2 x^2
			+ \beta_3 x^3 ,\label{y1}\\
	y_2(x) &=& \beta_1
			+ 2( \beta_1 + \beta_2 ) x
			+ ( \beta_2 + \beta_3 ) x^2 ,\\
	y_3(x) &=& \beta_1
			+ 2 \beta_2 x
			+ \beta_3 x^2 ,\\
	y_4(x) &=& - \beta_1
			+ (\beta_1 - 2\beta_2) x
			+ (2\beta_2 - \beta_3) x^2
			+ \beta_3 x^3 .
\eea
Now we can get the field equations by varying
action~\eqref{finalAction}
with respect to $N$ and $a$, respectively,
\bea
	& \disp\frac{1}{M_p^2} \rho
	= 2 T f_T - f + T - m^2 y_1(x),\label{matter_density}\\
	& \disp\frac{1}{M_p^2} P
	= -8 T f_{T T} \dot{H}
		- T + f - 2 T f_T
		- 4 \left( 1 + f_T \right) \dot{H}
		+ m^2 y_2(x) ,\label{matter_pressure}
\eea
where a dot denotes $\frac{d}{N d \tau}$,
$f_T \equiv df/d T$,
$f_{T T} \equiv d^2 f/d T^2$,
and $\rho$, $P$ are the energy density and pressure of
all perfect fluids of generic matter, respectively.
Notice that we do not need to know the explicit form of $S_m$
in the above derivation.
As is well known, the energy-momentum tensor of a perfect fluid
takes a diagonal form in the comoving coordinates, namely
\be
	\tensor{T}{^\mu_\nu} = \mathrm{diag}\left( -\rho, P, P, P \right).
\ee
And from the standard definition of energy-momentum tensor
\be
	T^{\mu \nu}
	\equiv \frac{2}{\sqrt{-g}} \frac{\delta S_m}{\delta g_{\mu \nu}},
\ee
we can easily read off $\rho$ and $P$ as
\be
	\rho = - \frac{1}{a^3} \frac{\delta S_m}{\delta N},
	\qquad P = \frac{1}{3 N a^2} \frac{\delta S_m}{\delta a}.
\ee
From Eqs.~\eqref{matter_density} and \eqref{matter_pressure},
we can get the corresponding energy conservation equation
for matter,
\be\label{matter_energy_conservation}
	\dot{\rho}
	= - 3 H ( \rho + P ) + 3 m^2 M_p^2 H y_3(x) .
\ee
We can recast Eqs.~\eqref{matter_density} and
\eqref{matter_pressure} in the similar form as in the
GR case by introducing the energy density and
pressure of the effective DE as
\bea
	&\disp \rho_{de} = M_p^2 \left[ f - 2 T f_T + m^2 y_1(x)
           \right] ,\label{DE_density}\\[1mm]
	&\disp P_{de} = M_p^2 \left\{
			\left( 8 T f_{T T} + 4 f_T \right) \dot{H}
			+ 2 T f_T - f
			- m^2 y_2(x)
			\right\} .\label{DE_pressure}
\eea
Then we get the modified Friedmann equations as
\bea
	& \disp 6 M_p^2 H^2 = \rho_{total} ,\\
	& \disp 4 M_p^2 \dot{H} = -\left( \rho_{total} + P_{total} \right) ,
\eea
in which $\rho_{total} = \rho + \rho_{de}$,
$P_{total} = P + P_{de}$.
From Eqs.~\eqref{DE_density} and~\eqref{DE_pressure},
one can obtain the corresponding energy conservation
equation for DE as
\be\label{DE_energy_conservation}
	\dot{\rho}_{de}
	= -3 H \left( \rho_{de} + P_{de} \right)
		- 3 m^2 M_p^2 H y_3(x).
\ee
Combining Eqs.~\eqref{matter_energy_conservation} and
\eqref{DE_energy_conservation} yields
\be\label{energy_conservation}
	\frac{d}{d t} \left( a^3 \rho_{total} \right) = -3 a^3 H p_{total},
\ee
which is the usual conservation equation for total energy. We
see that matter and DE are interacted through the graviton. We
define the equation-of-state (EoS) parameter of the effective DE as
\be\label{eos_de}
	w_{de}
	\equiv \frac{P_{de}}{\rho_{de}}
	= -1 + \frac{
			4 \left( 2 T f_{T T} + f_T \right) \dot{H}
			+ m^2 y_4(x)} {-2 T f_T + f + m^2 y_1(x)
			} .
\ee

%%%%%%%%%%%%%%%%%%%%%%%%%%%%%%%%%%%%%%%%%%%%%%%%%%%%%%%%%%%%%%%%%%%%%%
%%%%%%%%%%%%%%%%%%%%%%%%%%%%% section 4 %%%%%%%%%%%%%%%%%%%%%%%%%%%%%%
%%%%%%%%%%%%%%%%%%%%%%%%%%%%%%%%%%%%%%%%%%%%%%%%%%%%%%%%%%%%%%%%%%%%%%
\section{\label{observations}{Observational constraints}}

Here we are interested in the late-time behavior of $f(T)$
massive gravity. In this section, we study two specific functionals
of $f(T)$ as  concrete examples. These two examples will be
constrained by the recent cosmological data. We then study the
evolution of the effective EoS parameter of DE.

%%%%%%%%%%%%%%%%%%%%%%%%%%%%%%%%%%%%%%%%%%%%%%%%%%%%%%%%%%%%%%%%%%%%%%
\subsection{General set-up}

Since we are interested in the late-time universe, we ignore
the radiation component and only
consider the pressureless matter, whose energy density is $\rho_m$
and pressure is $P_m = 0$.
Furthermore, we also set $\beta_2 = \beta_3 = 0$,
since at the background level, the terms $3 \beta_2 m^2 a^{-2}$ and
$3 \beta_3 m^2 a^{-3}$ in Eq.~\eqref{DE_density}
are indistinguishable from the space curvature and pressureless
matter terms, respectively. Having these in mind, and by using
Eqs.~\eqref{y1} and \eqref{matter_pressure}, the effective
EoS parameter of DE~\eqref{eos_de} can be further simplified~to
\be
	w_{de}
	= - \frac{ 2 T f_{T T}
				- f_T + f/T
				+ \frac{T_0}{T} \Omega_{x0} ( 2 z/3 + 1 )
				}{ \left( 2 T f_{T T} + f_T + 1 \right) \left[
						-2 f_T + f/T +  \frac{T_0}{T} \Omega_{x0} (z+1)
						\right]
					} ,
\ee
where the fractional density of the
pressureless matter and the graviton are given by
\be
	\Omega_m = \frac{\rho_m}{6 M_p^2 H^2}
	\quad \mathrm{and} \quad
	\Omega_x = \frac{\beta_1 m^2}{2 H^2} ,
\ee
respectively. Note that we always use a subscript ``$0$'' to denote
the present value of corresponding quantity.
So $T_0 = T(z = 0)$ and $\Omega_{x0} = \Omega_x(z = 0)$, with
the redshift $z$ defined as $z \equiv a^{-1} - 1$.
Consequently, we find the expression for the dimensionless
Hubble parameter $E \equiv H/H_0$, namely
\be\label{generalE}
	E^2(z)
	= \Omega_{x0} ( 1 + z )
		+ \Omega_{m0} ( 1 + z )^3
		+ \frac{f}{T_0}
		- 2 E^2(z)\, f_T .
\ee
Note that $E$ appears in both sides of this equation, and
implicitly in $f$ and $f_T$.

We now briefly review the cosmological data and fitting methodology
used in constraining the model parameters.
We will perform a joint analysis of the Type Ia Supernovae (SNIa),
the baryonic acoustic oscillation (BAO) and the cosmic
microwave background (CMB) data to break the degeneracy between
the model parameters. For SNIa, we use the Union2.1 dataset
\cite{Suzuki:2011hu} which consists of 580 data points.
These data are given in terms of the distance  modulus $\mu_{obs}(z_i)$.
By definition, the theoretical distance modulus is given by
\be
 \mu_{th}(z_i)\equiv 5\log_{10}D_L(z_i)+\mu_0\,,
\ee
where $\mu_0 \equiv 42.38 - 5 \log_{10} h$ with $h$ the Hubble
constant $H_0$ in units of $100\,{\rm km/s/Mpc}$.
Here, the luminosity distance can be calculated as
\be
 D_L(z)=(1+z)\int_0^z \frac{d\tilde{z}}{E(\tilde{z};{\bf p})}\,,
\ee
in which ${\bf p}$ denotes the model parameters.
Consequently, the $\chi^2$ from 580 Union2.1 SNIa is given by
\be\label{eq3}
 \chi^2_{SN}({\bf p})=\sum\limits_{i}\frac{\left[
 \mu_{obs}(z_i)-\mu_{th}(z_i)\right]^2}{\sigma^2(z_i)}\,,
\ee
where $\sigma$ is the corresponding $1\sigma$ error.
Following \cite{Nesseris:2005ur}, we marginalize over $\mu_0$ by
expanding the $\chi^2_{SN}$ with respect to $\mu_0$ as
\be\label{eq4}
 \chi^2_{SN}({\bf p})=\tilde{A}-2\mu_0\tilde{B}+\mu_0^2\tilde{C}\,,
\ee
where
 $$\tilde{A}({\bf p})=\sum\limits_{i}\frac{\left[
 \mu_{obs}(z_i)-\mu_{th}(z_i;\mu_0=0,{\bf p})\right]^2}
 {\sigma_{\mu_{obs}}^2(z_i)}\,,$$
 $$\tilde{B}({\bf p})=\sum\limits_{i}\frac{\mu_{obs}(z_i)
 -\mu_{th}(z_i;\mu_0=0,{\bf p})}{\sigma_{\mu_{obs}}^2(z_i)}\,,
 ~~~~~~~~~~~
 \tilde{C}=\sum\limits_{i}\frac{1}{\sigma_{\mu_{obs}}^2(z_i)}\,.$$
Eq.~(\ref{eq4}) has a minimum for $\mu_0=\tilde{B}/\tilde{C}$ at
\be\label{eq5}
 \tilde{\chi}^2_{SN}({\bf p})=\tilde{A}({\bf p})
 -\frac{\tilde{B}({\bf p})^2}{\tilde{C}}\,.
\ee
Since $\chi^2_{SN,\,min}=\tilde{\chi}^2_{SN,\,min}$ (up to
a constant), we can instead minimize
$\tilde{\chi}^2_{SN}$ which is independent of $\mu_0$.
For the data of CMB and BAO, we use the shift parameter $R$
from CMB, and the distance parameter $A$ from the measurement
of the BAO peak in the distribution of SDSS luminous red
galaxies, since they are
model-independent and contain the main information of the
observations of CMB and BAO, respectively (see
e.g. \cite{Wang:2006ts}).
The shift parameter $R$ of CMB is defined by \cite{Bond:1997wr}
\be
 R\equiv\Omega_{m0}^{1/2}\int_0^{z_\ast}
 \frac{d\tilde{z}}{E(\tilde{z})}\,,
\ee
where the redshift of recombination $z_\ast$
is determined to be $1089.90$ by the Planck 2015
data \cite{Planck:2015xua}. On the other hand,
the Planck 2015 data have also determined the observed value
of shift parameter $R_{obs}$ to be
$1.7382 \pm 0.0088$~\cite{Ade:2015rim}. So, the $\chi^2$ for CMB is
\be
	\chi^2_{CMB}=\frac{(R-R_{obs})^2}{\sigma_R^2}.
\ee
The distance parameter $A$
of the measurement of the BAO peak in the distribution of SDSS
luminous red galaxies \cite{Eisenstein:2005su} is given by
\be
 A\equiv\Omega_{m0}^{1/2}E(z_b)^{-1/3}\left[\frac{1}{z_b}
 \int_0^{z_b}\frac{d\tilde{z}}{E(\tilde{z})}\right]^{2/3},
\ee
where $z_b=0.35$. In \cite{Eisenstein:2005su}, the value of $A$ has
been determined to be $0.469\,(n_s/0.98)^{-0.35}\pm 0.017$.
Here the scalar spectral index $n_s$ is taken to be $0.9741$ by the
Planck 2015 data \cite{Ade:2015rim}. And the corresponding
$\chi^2$ for BAO is
\be
	\chi^2_{BAO}=\frac{(A-A_{obs})^2}{\sigma_A^2}.
\ee
So, the total $\chi^2$ is given by
\be\label{eq8}
 \chi^2=\tilde{\chi}^2_{SN}+\chi^2_{CMB}+\chi^2_{BAO}\,.
\ee
Then we can minimize the total $\chi^2$ to get
the best-fit values of model parameters. The
 $68.3\%$ and $95.4\%$ confidence levels are determined
 by $\Delta\chi^2\equiv\chi^2-\chi^2_{min}\leq 3.53$ and
 $8.02$, respectively, if there are 3 free model parameters.

\begin{figure}[htbp!]
\centering
	\includegraphics[width = \textwidth]{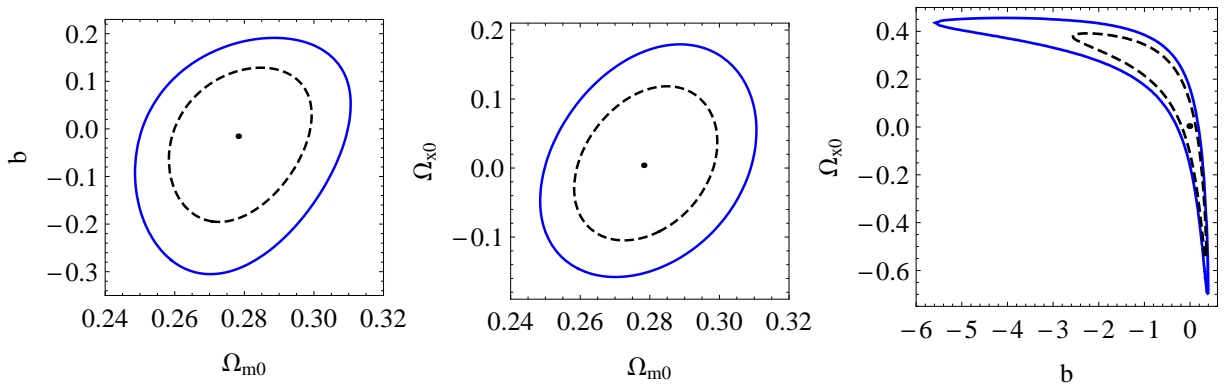}
	\caption{\label{contours_f1MG}
	(Color online) The $68.3\%$ (dashed) and $95.4\%$ (solid) confidence
	level contours for the power-law $f(T)$ massive gravity
	in the $(\Omega_{m0},\, b)$ plane (left),
	the $(\Omega_{m0},\, \Omega_{x0})$ plane (middle)
	and the $(b,\, \Omega_{x0})$ plane (right).
	The best-fit parameters are indicated by the solid points.
	}
\end{figure}

\vspace{-5mm} % used here just for a better typesetting

%%%%%%%%%%%%%%%%%%%%%%%%%%%%%%%%%%%%%%%%%%%%%%%%%%%%%%%%%%%%%%%%%%%%%%
\subsection{The power-law case}

\begin{figure}[htbp!]
\centering
	\includegraphics[width = \textwidth]{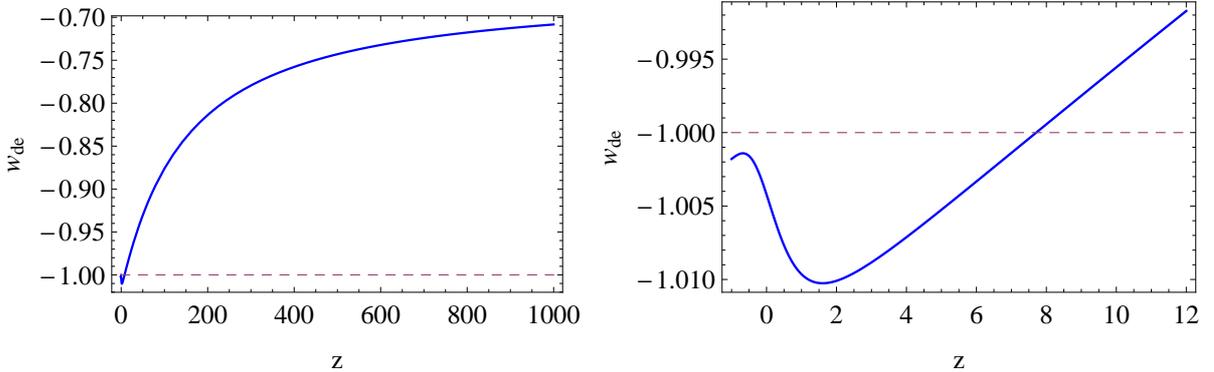}
	\caption{\label{wde_f1MG}
	(Color online) The evolutionary curve of the effective
	EoS parameter of DE with the best-fit values of $\Omega_{m0}$,
	$\Omega_{x0}$ and $b$, for the power-law $f(T)$ massive gravity.
	The left and right panels correspond to large
	and small range of redshift, respectively.
	The phantom divide line ($w_{de} = -1$) is also indicated by
	the dashed lines.
	}
\end{figure}

In this subsection, we consider the power-law functional of $f(T)$
first introduced by Bengochea \textit{et al.} in
\cite{Bengochea:2008gz} (see also \cite{Linder:2010py}), which reads
\be
	f(T) = \alpha T^b ,
\ee
where $\alpha$ and $b$ are both constants.
Demanding Eq.~\eqref{generalE} to be satisfied at
redshift $z = 0$, we have
\be\label{alpha0}
	\alpha
	= \frac{1 - \Omega_{x0} - \Omega_{m0}}{1-2b}\left(6 H_0^2\right)^{1-b}.
\ee
Substituting Eq.~\eqref{alpha0} back into Eq.~\eqref{generalE},
the corresponding background evolution reads
\be
	E^2
	= \Omega_{x0} ( 1 + z )
		+ \Omega_{m0} ( 1 + z )^3
		+ \left( 1 - \Omega_{x0} - \Omega_{m0} \right) E^{2 b} .
\ee
We note that, when $\Omega_{x0} = b = 0$, this model corresponds to
$\Lambda$CDM model in fact.
There are 3 free parameters in this model, namely $\Omega_{x0}$,
$\Omega_{m0}$, and $b$.
By minimizing the corresponding total $\chi^2$ in Eq.~(\ref{eq8}),
we find the best-fit parameters $\Omega_{x0} = 0.004$,
$\Omega_{m0} = 0.278$, and $b = -0.015$,
while $\chi^2_{min} = 562.255$.
In Fig.~\ref{contours_f1MG}, we present the corresponding
$68.3\%$ and $95.4\%$ confidence level contours
for the power-law $f(T)$ massive gravity
in the $(\Omega_{m0},\, b)$, the $(\Omega_{m0},\, \Omega_{x0})$
and the $(b,\, \Omega_{x0})$ planes, respectively.
We find that $\Lambda$CDM model
(corresponding to $\Omega_{x0} = b = 0$) is still consistent
with the observations at the $68.3\%$ confidence level.
In Fig.~\ref{wde_f1MG}, we present the evolutionary curve
of $w_{de}$ with the best-fit values of model parameters.
Apparently, the power-law case shows a phantom crossing
behavior with the crossing of phantom divide line occurring
at redshift $z \simeq 7.71$.

\vspace{-2mm}  % used here just for a better typesetting

%%%%%%%%%%%%%%%%%%%%%%%%%%%%%%%%%%%%%%%%%%%%%%%%%%%%%%%%%%%%%%%%%%%%%%
\subsection{The exponential case}

\begin{figure}[tb]
\centering
	\includegraphics[width = \textwidth]{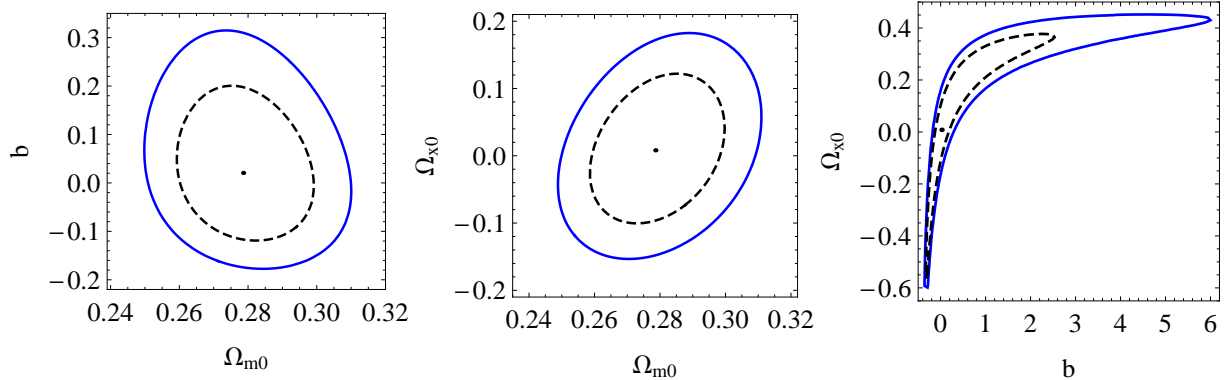}
	\caption{\label{contours_f2MG}
	(Color online) The $68.3\%$ (dashed) and $95.4\%$ (solid) confidence
	level contours for the exponential $f(T)$ massive gravity
	in the $(\Omega_{m0},\, b)$ plane (left),
	the $(\Omega_{m0},\, \Omega_{x0})$ plane (middle)
	and the $(b,\, \Omega_{x0})$ plane (right).
	The best-fit parameters are indicated by the solid points.
	}
\end{figure}

\begin{figure}[htb]
\centering
	\includegraphics[width = \textwidth]{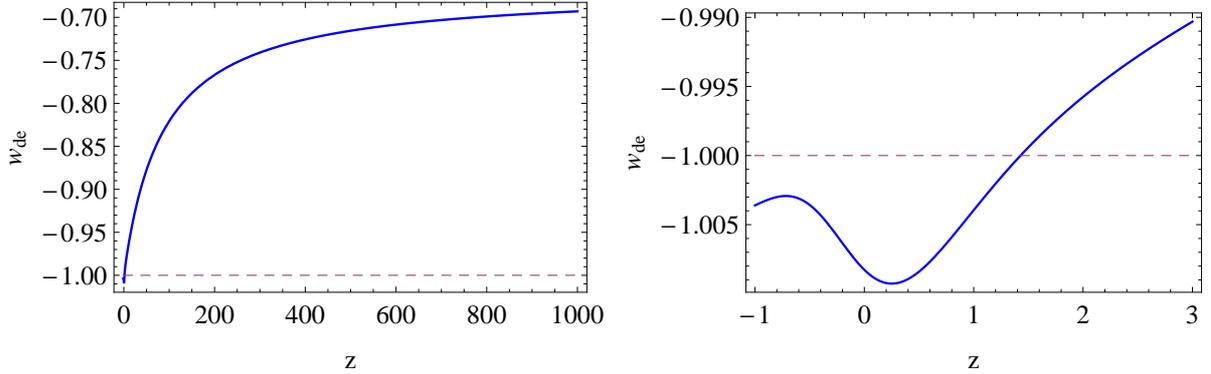}
	\caption{\label{wde_f2MG}
	(Color online) The evolutionary curve of the effective
	EoS parameter of DE with the best-fit values of $\Omega_{m0}$,
	$\Omega_{x0}$ and $b$, for the exponential $f(T)$ massive gravity.
	The left and right panels correspond to large
	and small range of redshift, respectively.
	The phantom divide line ($w_{de} = -1$) is also indicated by
	the dashed lines.
	}
\end{figure}

In this subsection, we consider the exponential functional of $f(T)$
first introduced by Linder in \cite{Linder:2010py}, which reads
\be
	f(T) = -\alpha T\left( 1 - e^{b \frac{T_0}{T}} \right) ,
\ee
where $\alpha$ and $b$ are both constants.
Demanding Eq.~\eqref{generalE} to be satisfied at
redshift $z = 0$, we have
\be\label{alpha}
	\alpha = \frac{1 - \Omega_{m0} -\Omega_{x0}}{ 1 - ( 1 - 2 b ) e^b} .
\ee
Substituting Eq.~\eqref{alpha} back into Eq.~\eqref{generalE},
the corresponding background evolution reads
\be
	E^2
	= \Omega_{x0} ( 1 + z )
		+ \Omega_{m0} ( 1 + z )^3
		+ \frac{1 - \Omega_{m0} -\Omega_{x0}}{ 1 - ( 1 - 2 b ) e^b}
		    E^2 \left[
		    	1
		    	- e^{\frac{b}{E^2}} \left( 1 - 2 \frac{b}{E^2} \right)
		    	\right] .
\ee
We note that, when $\Omega_{x0} = b = 0$, this model corresponds to
$\Lambda$CDM model in fact.
There are 3 free parameters in this model, namely $\Omega_{x0}$,
$\Omega_{m0}$, and $b$.
By minimizing the corresponding total $\chi^2$ in Eq.~(\ref{eq8}),
we find the best-fit parameters $\Omega_{x0} = 0.008$,
$\Omega_{m0} = 0.279$, and $b = 0.021$,
while $\chi^2_{min} = 562.262$.
In Fig.~\ref{contours_f2MG}, we present the corresponding
$68.3\%$ and $95.4\%$ confidence level contours
for the exponential $f(T)$ massive gravity
in the $(\Omega_{m0},\, b)$, the $(\Omega_{m0},\, \Omega_{x0})$
and the $(b,\, \Omega_{x0})$ planes, respectively.
We find that $\Lambda$CDM model
(corresponding to $\Omega_{x0} = b = 0$) is still consistent
with the observations at the $68.3\%$ confidence level.
In Fig.~\ref{wde_f2MG}, we present the evolutionary curve
of $w_{de}$ with the best-fit values of model parameters.
Apparently, the exponential case shows a phantom crossing
behavior  with the crossing of phantom divide line occurring
at redshift $z \simeq 1.43$.

\vspace{-2.5mm}  % used here just for a better typesetting

%%%%%%%%%%%%%%%%%%%%%%%%%%%%%%%%%%%%%%%%%%%%%%%%%%%%%%%%%%%%%%%%%%%%%%
%%%%%%%%%%%%%%%%%%%%%%%%%%%%% section 5 %%%%%%%%%%%%%%%%%%%%%%%%%%%%%%
%%%%%%%%%%%%%%%%%%%%%%%%%%%%%%%%%%%%%%%%%%%%%%%%%%%%%%%%%%%%%%%%%%%%%%
\section{\label{Conclusions}{Conclusions}}

In this paper, we extend $f(T)$ theory and dRGT massive gravity to
a new kind of modified gravity model, namely $f(T)$ non-linear massive
gravity, by adding the dRGT mass term to $f(T)$ theory.
This mass term is formulated in the vierbein formalism to agree
with the teleparallel space-time.
Since the resulting field equations are $2$nd order, and the mass
term does not contain the square root of the metric when using
vierbein formalism, this theory is easier to analyse than
$f(R)$ non-linear massive gravity. Besides, thanks to the
rich structure of $f(T)$ sector and massive graviton,
it is natural to expect that this theory could also unify the early
inflation and late-time acceleration in a consistent framework, and
we leave this issue to the future works.

We then investigate the cosmological evolution of $f(T)$
non-linear massive gravity.
In particular, we study the power-law and exponential cases of
$f(T)$ massive gravity
as two toy models.
We then perform a joint constraint on the model parameters
by the recent data of SNIa, CMB and BAO.
We find that the power-law and exponential $f(T)$ massive gravity
are consistent with these cosmological observations.
Furthermore, we explore the evolution of the effective EoS parameter
of DE, and find that it can realize the crossing of the phantom
divide line from the quintessence-like phase ($w_{de} > -1$) to the
phantom-like
one ($w_{de} < -1$) by using the best-fit parameters obtained
from the above cosmological constraints. We note here that the
recent data shows great possibility that the EoS parameter of
DE crosses the phantom divide line from the quintessence-like phase
to the phantom-like phase as the redshift $z$ decreases in the near
past \cite{Jassal:2006gf,Alam:2004jy,Nesseris:2006er,Wu:2006bb}.
Although there exist some complicated specific $f(T)$ models to
realize the phantom crossing behavior \cite{Wu:2010av,Bamba:2010wb},
in general, especially in the original power-law
and exponential $f(T)$ gravity, phantom crossing
is impossible \cite{Wu:2010mn,Bamba:2010wb,Bamba:2010iw}. So,
our results are of interest.

%%%%%%%%%%%%%%%%%%%%%%%%%%%%%%%%%%%%%%%%%%%%%%%%%%%%%%%%%%%%%%%%%%%%%%
%%%%%%%%%%%%%%%%%%%%%%%%%% acknowledgements %%%%%%%%%%%%%%%%%%%%%%%%%%
%%%%%%%%%%%%%%%%%%%%%%%%%%%%%%%%%%%%%%%%%%%%%%%%%%%%%%%%%%%%%%%%%%%%%%
\begin{acknowledgments}
We thank Savvas Nesseris for helpful discussion and
providing his Mathematica code for data constraint, which
greatly improves our work. We are grateful to Jing Liu,
Xiao-Peng~Yan, Ya-Nan~Zhou, Xiao-Bo~Zou, and Hong-Yu~Li
for kind help and discussions.
This work was supported in part by NSFC under
Grants No.~11175016 and No.~10905005, as well as NCET under
Grant No.~NCET-11-0790.
\end{acknowledgments}

%%%%%%%%%%%%%%%%%%%%%%%%%%%%%%%%%%%%%%%%%%%%%%%%%%%%%%%%%%%%%%%%%%%%%%
%%%%%%%%%%%%%%%%%%%%%%%%%%%%% references %%%%%%%%%%%%%%%%%%%%%%%%%%%%%
%%%%%%%%%%%%%%%%%%%%%%%%%%%%%%%%%%%%%%%%%%%%%%%%%%%%%%%%%%%%%%%%%%%%%%
\bibliography{./fnmg}

%merlin.mbs apsrev4-1.bst 2010-07-25 4.21a (PWD, AO, DPC) hacked
%Control: key (0)
%Control: author (8) initials jnrlst
%Control: editor formatted (1) identically to author
%Control: production of article title (-1) disabled
%Control: page (0) single
%Control: year (1) truncated
%Control: production of eprint (0) enabled
\begin{thebibliography}{90}%
\makeatletter
\providecommand \@ifxundefined [1]{%
 \@ifx{#1\undefined}
}%
\providecommand \@ifnum [1]{%
 \ifnum #1\expandafter \@firstoftwo
 \else \expandafter \@secondoftwo
 \fi
}%
\providecommand \@ifx [1]{%
 \ifx #1\expandafter \@firstoftwo
 \else \expandafter \@secondoftwo
 \fi
}%
\providecommand \natexlab [1]{#1}%
\providecommand \enquote  [1]{``#1''}%
\providecommand \bibnamefont  [1]{#1}%
\providecommand \bibfnamefont [1]{#1}%
\providecommand \citenamefont [1]{#1}%
\providecommand \href@noop [0]{\@secondoftwo}%
\providecommand \href [0]{\begingroup \@sanitize@url \@href}%
\providecommand \@href[1]{\@@startlink{#1}\@@href}%
\providecommand \@@href[1]{\endgroup#1\@@endlink}%
\providecommand \@sanitize@url [0]{\catcode `\\12\catcode `\$12\catcode
  `\&12\catcode `\#12\catcode `\^12\catcode `\_12\catcode `\%12\relax}%
\providecommand \@@startlink[1]{}%
\providecommand \@@endlink[0]{}%
\providecommand \url  [0]{\begingroup\@sanitize@url \@url }%
\providecommand \@url [1]{\endgroup\@href {#1}{\urlprefix }}%
\providecommand \urlprefix  [0]{URL }%
\providecommand \Eprint [0]{\href }%
\providecommand \doibase [0]{http://dx.doi.org/}%
\providecommand \selectlanguage [0]{\@gobble}%
\providecommand \bibinfo  [0]{\@secondoftwo}%
\providecommand \bibfield  [0]{\@secondoftwo}%
\providecommand \translation [1]{[#1]}%
\providecommand \BibitemOpen [0]{}%
\providecommand \bibitemStop [0]{}%
\providecommand \bibitemNoStop [0]{.\EOS\space}%
\providecommand \EOS [0]{\spacefactor3000\relax}%
\providecommand \BibitemShut  [1]{\csname bibitem#1\endcsname}%
\let\auto@bib@innerbib\@empty
%</preamble>
\bibitem [{\citenamefont {Zwicky}(1933)}]{Zwicky:1933gu}%
  \BibitemOpen
  \bibfield  {author} {\bibinfo {author} {\bibfnamefont {F.}~\bibnamefont
  {Zwicky}},\ }\href@noop {} {\bibfield  {journal} {\bibinfo  {journal}
  {Helv.Phys.Acta}\ }\textbf {\bibinfo {volume} {6}},\ \bibinfo {pages} {110}
  (\bibinfo {year} {1933})}\BibitemShut {NoStop}%
%%CITATION = HPACA,6,110;%%
\bibitem [{\citenamefont {{Zwicky}}(1937)}]{1937ApJ}%
  \BibitemOpen
  \bibfield  {author} {\bibinfo {author} {\bibfnamefont {F.}~\bibnamefont
  {{Zwicky}}},\ }\href {\doibase 10.1086/143864} {\bibfield  {journal}
  {\bibinfo  {journal} {\apj}\ }\textbf {\bibinfo {volume} {86}},\ \bibinfo
  {pages} {217} (\bibinfo {year} {1937})}\BibitemShut {NoStop}%
\bibitem [{\citenamefont {Riess}\ \emph {et~al.}(1998)\citenamefont {Riess}
  \emph {et~al.}}]{Riess:1998cb}%
  \BibitemOpen
  \bibfield  {author} {\bibinfo {author} {\bibfnamefont {A.~G.}\ \bibnamefont
  {Riess}} \emph {et~al.} (\bibinfo {collaboration} {Supernova Search Team}),\
  }\href {\doibase 10.1086/300499} {\bibfield  {journal} {\bibinfo  {journal}
  {Astron.J.}\ }\textbf {\bibinfo {volume} {116}},\ \bibinfo {pages} {1009}
  (\bibinfo {year} {1998})},\ \Eprint {http://arxiv.org/abs/astro-ph/9805201}
  {astro-ph/9805201} \BibitemShut {NoStop}%
%%CITATION = ASTRO-PH/9805201;%%
\bibitem [{\citenamefont {Perlmutter}\ \emph {et~al.}(1999)\citenamefont
  {Perlmutter} \emph {et~al.}}]{Perlmutter:1998np}%
  \BibitemOpen
  \bibfield  {author} {\bibinfo {author} {\bibfnamefont {S.}~\bibnamefont
  {Perlmutter}} \emph {et~al.} (\bibinfo {collaboration} {Supernova Cosmology
  Project}),\ }\href {\doibase 10.1086/307221} {\bibfield  {journal} {\bibinfo
  {journal} {Astrophys.J.}\ }\textbf {\bibinfo {volume} {517}},\ \bibinfo
  {pages} {565} (\bibinfo {year} {1999})},\ \Eprint
  {http://arxiv.org/abs/astro-ph/9812133} {astro-ph/9812133} \BibitemShut
  {NoStop}%
%%CITATION = ASTRO-PH/9812133;%%
\bibitem [{\citenamefont {Spergel}\ \emph {et~al.}(2003)\citenamefont {Spergel}
  \emph {et~al.}}]{Spergel:2003cb}%
  \BibitemOpen
  \bibfield  {author} {\bibinfo {author} {\bibfnamefont {D.}~\bibnamefont
  {Spergel}} \emph {et~al.} (\bibinfo {collaboration} {WMAP Collaboration}),\
  }\href {\doibase 10.1086/377226} {\bibfield  {journal} {\bibinfo  {journal}
  {Astrophys.J.Suppl.}\ }\textbf {\bibinfo {volume} {148}},\ \bibinfo {pages}
  {175} (\bibinfo {year} {2003})},\ \Eprint
  {http://arxiv.org/abs/astro-ph/0302209} {astro-ph/0302209} \BibitemShut
  {NoStop}%
%%CITATION = ASTRO-PH/0302209;%%
\bibitem [{\citenamefont {Tegmark}\ \emph {et~al.}(2004)\citenamefont {Tegmark}
  \emph {et~al.}}]{Tegmark:2003ud}%
  \BibitemOpen
  \bibfield  {author} {\bibinfo {author} {\bibfnamefont {M.}~\bibnamefont
  {Tegmark}} \emph {et~al.} (\bibinfo {collaboration} {SDSS Collaboration}),\
  }\href {\doibase 10.1103/PhysRevD.69.103501} {\bibfield  {journal} {\bibinfo
  {journal} {Phys.Rev.}\ }\textbf {\bibinfo {volume} {D69}},\ \bibinfo {pages}
  {103501} (\bibinfo {year} {2004})},\ \Eprint
  {http://arxiv.org/abs/astro-ph/0310723} {astro-ph/0310723} \BibitemShut
  {NoStop}%
%%CITATION = ASTRO-PH/0310723;%%
\bibitem [{\citenamefont {Allen}\ \emph {et~al.}(2004)\citenamefont {Allen}
  \emph {et~al.}}]{Allen:2004cd}%
  \BibitemOpen
  \bibfield  {author} {\bibinfo {author} {\bibfnamefont {S.}~\bibnamefont
  {Allen}} \emph {et~al.},\ }\href {\doibase 10.1111/j.1365-2966.2004.08080.x}
  {\bibfield  {journal} {\bibinfo  {journal} {Mon.Not.Roy.Astron.Soc.}\
  }\textbf {\bibinfo {volume} {353}},\ \bibinfo {pages} {457} (\bibinfo {year}
  {2004})},\ \Eprint {http://arxiv.org/abs/astro-ph/0405340} {astro-ph/0405340}
  \BibitemShut {NoStop}%
%%CITATION = ASTRO-PH/0405340;%%
\bibitem [{\citenamefont {Riess}\ \emph {et~al.}(2004)\citenamefont {Riess}
  \emph {et~al.}}]{Riess:2004nr}%
  \BibitemOpen
  \bibfield  {author} {\bibinfo {author} {\bibfnamefont {A.~G.}\ \bibnamefont
  {Riess}} \emph {et~al.} (\bibinfo {collaboration} {Supernova Search Team}),\
  }\href {\doibase 10.1086/383612} {\bibfield  {journal} {\bibinfo  {journal}
  {Astrophys.J.}\ }\textbf {\bibinfo {volume} {607}},\ \bibinfo {pages} {665}
  (\bibinfo {year} {2004})},\ \Eprint {http://arxiv.org/abs/astro-ph/0402512}
  {astro-ph/0402512} \BibitemShut {NoStop}%
%%CITATION = ASTRO-PH/0402512;%%
\bibitem [{\citenamefont {Spergel}\ \emph {et~al.}(2007)\citenamefont {Spergel}
  \emph {et~al.}}]{Spergel:2006hy}%
  \BibitemOpen
  \bibfield  {author} {\bibinfo {author} {\bibfnamefont {D.}~\bibnamefont
  {Spergel}} \emph {et~al.} (\bibinfo {collaboration} {WMAP Collaboration}),\
  }\href {\doibase 10.1086/513700} {\bibfield  {journal} {\bibinfo  {journal}
  {Astrophys.J.Suppl.}\ }\textbf {\bibinfo {volume} {170}},\ \bibinfo {pages}
  {377} (\bibinfo {year} {2007})},\ \Eprint
  {http://arxiv.org/abs/astro-ph/0603449} {astro-ph/0603449} \BibitemShut
  {NoStop}%
%%CITATION = ASTRO-PH/0603449;%%
\bibitem [{\citenamefont {Ade}\ \emph {et~al.}(2014{\natexlab{a}})\citenamefont
  {Ade} \emph {et~al.}}]{Ade:2013ktc}%
  \BibitemOpen
  \bibfield  {author} {\bibinfo {author} {\bibfnamefont {P.}~\bibnamefont
  {Ade}} \emph {et~al.} (\bibinfo {collaboration} {Planck Collaboration}),\
  }\href {\doibase 10.1051/0004-6361/201321529} {\bibfield  {journal} {\bibinfo
   {journal} {Astron.Astrophys.}\ }\textbf {\bibinfo {volume} {571}},\ \bibinfo
  {pages} {A1} (\bibinfo {year} {2014}{\natexlab{a}})},\ \Eprint
  {http://arxiv.org/abs/1303.5062} {arXiv:1303.5062 [astro-ph.CO]} \BibitemShut
  {NoStop}%
%%CITATION = ARXIV:1303.5062;%%
\bibitem [{\citenamefont {Ade}\ \emph {et~al.}(2014{\natexlab{b}})\citenamefont
  {Ade} \emph {et~al.}}]{Ade:2013zuv}%
  \BibitemOpen
  \bibfield  {author} {\bibinfo {author} {\bibfnamefont {P.}~\bibnamefont
  {Ade}} \emph {et~al.} (\bibinfo {collaboration} {Planck Collaboration}),\
  }\href {\doibase 10.1051/0004-6361/201321591} {\bibfield  {journal} {\bibinfo
   {journal} {Astron.Astrophys.}\ }\textbf {\bibinfo {volume} {571}},\ \bibinfo
  {pages} {A16} (\bibinfo {year} {2014}{\natexlab{b}})},\ \Eprint
  {http://arxiv.org/abs/1303.5076} {arXiv:1303.5076 [astro-ph.CO]} \BibitemShut
  {NoStop}%
%%CITATION = ARXIV:1303.5076;%%
\bibitem [{\citenamefont {Steffen}(2009)}]{Steffen:2008qp}%
  \BibitemOpen
  \bibfield  {author} {\bibinfo {author} {\bibfnamefont {F.~D.}\ \bibnamefont
  {Steffen}},\ }\href {\doibase 10.1140/epjc/s10052-008-0830-0} {\bibfield
  {journal} {\bibinfo  {journal} {Eur.Phys.J.}\ }\textbf {\bibinfo {volume}
  {C59}},\ \bibinfo {pages} {557} (\bibinfo {year} {2009})},\ \Eprint
  {http://arxiv.org/abs/0811.3347} {arXiv:0811.3347 [hep-ph]} \BibitemShut
  {NoStop}%
%%CITATION = ARXIV:0811.3347;%%
\bibitem [{\citenamefont {Bergstr{\"o}m}(2009)}]{Bergstrom:2009ib}%
  \BibitemOpen
  \bibfield  {author} {\bibinfo {author} {\bibfnamefont {L.}~\bibnamefont
  {Bergstr{\"o}m}},\ }\href {\doibase 10.1088/1367-2630/11/10/105006}
  {\bibfield  {journal} {\bibinfo  {journal} {New J.Phys.}\ }\textbf {\bibinfo
  {volume} {11}},\ \bibinfo {pages} {105006} (\bibinfo {year} {2009})},\
  \Eprint {http://arxiv.org/abs/0903.4849} {arXiv:0903.4849 [hep-ph]}
  \BibitemShut {NoStop}%
%%CITATION = ARXIV:0903.4849;%%
\bibitem [{\citenamefont {Feng}(2010)}]{Feng:2010gw}%
  \BibitemOpen
  \bibfield  {author} {\bibinfo {author} {\bibfnamefont {J.~L.}\ \bibnamefont
  {Feng}},\ }\href {\doibase 10.1146/annurev-astro-082708-101659} {\bibfield
  {journal} {\bibinfo  {journal} {Ann.Rev.Astron.Astrophys.}\ }\textbf
  {\bibinfo {volume} {48}},\ \bibinfo {pages} {495} (\bibinfo {year} {2010})},\
  \Eprint {http://arxiv.org/abs/1003.0904} {arXiv:1003.0904 [astro-ph.CO]}
  \BibitemShut {NoStop}%
%%CITATION = ARXIV:1003.0904;%%
\bibitem [{\citenamefont {Weinberg}(1989)}]{Weinberg:1988cp}%
  \BibitemOpen
  \bibfield  {author} {\bibinfo {author} {\bibfnamefont {S.}~\bibnamefont
  {Weinberg}},\ }\href {\doibase 10.1103/RevModPhys.61.1} {\bibfield  {journal}
  {\bibinfo  {journal} {Rev.Mod.Phys.}\ }\textbf {\bibinfo {volume} {61}},\
  \bibinfo {pages} {1} (\bibinfo {year} {1989})}\BibitemShut {NoStop}%
%%CITATION = RMPHA,61,1;%%
\bibitem [{\citenamefont {Fitch}\ \emph {et~al.}(1997)\citenamefont {Fitch}
  \emph {et~al.}}]{fitch1997critical}%
  \BibitemOpen
  \bibinfo {editor} {\bibfnamefont {V.~L.}\ \bibnamefont {Fitch}} \emph
  {et~al.},\ eds.,\ \href@noop {} {\emph {\bibinfo {title} {{Critical Problems
  in Physics}}}}\ (\bibinfo  {publisher} {Princeton University Press},\
  \bibinfo {year} {1997})\BibitemShut {NoStop}%
\bibitem [{\citenamefont {Peebles}\ and\ \citenamefont
  {Ratra}(2003)}]{Peebles:2002gy}%
  \BibitemOpen
  \bibfield  {author} {\bibinfo {author} {\bibfnamefont {P.}~\bibnamefont
  {Peebles}}\ and\ \bibinfo {author} {\bibfnamefont {B.}~\bibnamefont
  {Ratra}},\ }\href {\doibase 10.1103/RevModPhys.75.559} {\bibfield  {journal}
  {\bibinfo  {journal} {Rev.Mod.Phys.}\ }\textbf {\bibinfo {volume} {75}},\
  \bibinfo {pages} {559} (\bibinfo {year} {2003})},\ \Eprint
  {http://arxiv.org/abs/astro-ph/0207347} {astro-ph/0207347} \BibitemShut
  {NoStop}%
%%CITATION = ASTRO-PH/0207347;%%
\bibitem [{\citenamefont {Copeland}\ \emph {et~al.}(2006)\citenamefont
  {Copeland}, \citenamefont {Sami},\ and\ \citenamefont
  {Tsujikawa}}]{Copeland:2006wr}%
  \BibitemOpen
  \bibfield  {author} {\bibinfo {author} {\bibfnamefont {E.~J.}\ \bibnamefont
  {Copeland}}, \bibinfo {author} {\bibfnamefont {M.}~\bibnamefont {Sami}},\
  and\ \bibinfo {author} {\bibfnamefont {S.}~\bibnamefont {Tsujikawa}},\ }\href
  {\doibase 10.1142/S021827180600942X} {\bibfield  {journal} {\bibinfo
  {journal} {Int.J.Mod.Phys.}\ }\textbf {\bibinfo {volume} {D15}},\ \bibinfo
  {pages} {1753} (\bibinfo {year} {2006})},\ \Eprint
  {http://arxiv.org/abs/hep-th/0603057} {hep-th/0603057} \BibitemShut {NoStop}%
%%CITATION = HEP-TH/0603057;%%
\bibitem [{\citenamefont {'t~Hooft}\ and\ \citenamefont
  {Veltman}(1974)}]{tHooft:1974bx}%
  \BibitemOpen
  \bibfield  {author} {\bibinfo {author} {\bibfnamefont {G.}~\bibnamefont
  {'t~Hooft}}\ and\ \bibinfo {author} {\bibfnamefont {M.}~\bibnamefont
  {Veltman}},\ }\href@noop {} {\bibfield  {journal} {\bibinfo  {journal}
  {Annales Poincare Phys.Theor.}\ }\textbf {\bibinfo {volume} {A20}},\ \bibinfo
  {pages} {69} (\bibinfo {year} {1974})}\BibitemShut {NoStop}%
%%CITATION = AHPAA,A20,69;%%
\bibitem [{\citenamefont {Goroff}\ and\ \citenamefont
  {Sagnotti}(1986)}]{Goroff:1985th}%
  \BibitemOpen
  \bibfield  {author} {\bibinfo {author} {\bibfnamefont {M.~H.}\ \bibnamefont
  {Goroff}}\ and\ \bibinfo {author} {\bibfnamefont {A.}~\bibnamefont
  {Sagnotti}},\ }\href {\doibase 10.1016/0550-3213(86)90193-8} {\bibfield
  {journal} {\bibinfo  {journal} {Nucl.Phys.}\ }\textbf {\bibinfo {volume}
  {B266}},\ \bibinfo {pages} {709} (\bibinfo {year} {1986})}\BibitemShut
  {NoStop}%
%%CITATION = NUPHA,B266,709;%%
\bibitem [{\citenamefont {Sotiriou}\ and\ \citenamefont
  {Faraoni}(2010)}]{Sotiriou:2008rp}%
  \BibitemOpen
  \bibfield  {author} {\bibinfo {author} {\bibfnamefont {T.~P.}\ \bibnamefont
  {Sotiriou}}\ and\ \bibinfo {author} {\bibfnamefont {V.}~\bibnamefont
  {Faraoni}},\ }\href {\doibase 10.1103/RevModPhys.82.451} {\bibfield
  {journal} {\bibinfo  {journal} {Rev.Mod.Phys.}\ }\textbf {\bibinfo {volume}
  {82}},\ \bibinfo {pages} {451} (\bibinfo {year} {2010})},\ \Eprint
  {http://arxiv.org/abs/0805.1726} {arXiv:0805.1726 [gr-qc]} \BibitemShut
  {NoStop}%
%%CITATION = ARXIV:0805.1726;%%
\bibitem [{\citenamefont {De~Felice}\ and\ \citenamefont
  {Tsujikawa}(2010)}]{DeFelice:2010aj}%
  \BibitemOpen
  \bibfield  {author} {\bibinfo {author} {\bibfnamefont {A.}~\bibnamefont
  {De~Felice}}\ and\ \bibinfo {author} {\bibfnamefont {S.}~\bibnamefont
  {Tsujikawa}},\ }\href {\doibase 10.12942/lrr-2010-3} {\bibfield  {journal}
  {\bibinfo  {journal} {Living Rev.Rel.}\ }\textbf {\bibinfo {volume} {13}},\
  \bibinfo {pages} {3} (\bibinfo {year} {2010})},\ \Eprint
  {http://arxiv.org/abs/1002.4928} {arXiv:1002.4928 [gr-qc]} \BibitemShut
  {NoStop}%
%%CITATION = ARXIV:1002.4928;%%
\bibitem [{\citenamefont {Amendola}\ and\ \citenamefont
  {Tsujikawa}(2008)}]{Amendola:2007nt}%
  \BibitemOpen
  \bibfield  {author} {\bibinfo {author} {\bibfnamefont {L.}~\bibnamefont
  {Amendola}}\ and\ \bibinfo {author} {\bibfnamefont {S.}~\bibnamefont
  {Tsujikawa}},\ }\href {\doibase 10.1016/j.physletb.2007.12.041} {\bibfield
  {journal} {\bibinfo  {journal} {Phys.Lett.}\ }\textbf {\bibinfo {volume}
  {B660}},\ \bibinfo {pages} {125} (\bibinfo {year} {2008})},\ \Eprint
  {http://arxiv.org/abs/0705.0396} {arXiv:0705.0396 [astro-ph]} \BibitemShut
  {NoStop}%
%%CITATION = ARXIV:0705.0396;%%
\bibitem [{\citenamefont {Appleby}\ and\ \citenamefont
  {Battye}(2007)}]{Appleby:2007vb}%
  \BibitemOpen
  \bibfield  {author} {\bibinfo {author} {\bibfnamefont {S.~A.}\ \bibnamefont
  {Appleby}}\ and\ \bibinfo {author} {\bibfnamefont {R.~A.}\ \bibnamefont
  {Battye}},\ }\href {\doibase 10.1016/j.physletb.2007.08.037} {\bibfield
  {journal} {\bibinfo  {journal} {Phys.Lett.}\ }\textbf {\bibinfo {volume}
  {B654}},\ \bibinfo {pages} {7} (\bibinfo {year} {2007})},\ \Eprint
  {http://arxiv.org/abs/0705.3199} {arXiv:0705.3199 [astro-ph]} \BibitemShut
  {NoStop}%
%%CITATION = ARXIV:0705.3199;%%
\bibitem [{\citenamefont {Starobinsky}(2007)}]{Starobinsky:2007hu}%
  \BibitemOpen
  \bibfield  {author} {\bibinfo {author} {\bibfnamefont {A.~A.}\ \bibnamefont
  {Starobinsky}},\ }\href {\doibase 10.1134/S0021364007150027} {\bibfield
  {journal} {\bibinfo  {journal} {JETP Lett.}\ }\textbf {\bibinfo {volume}
  {86}},\ \bibinfo {pages} {157} (\bibinfo {year} {2007})},\ \Eprint
  {http://arxiv.org/abs/0706.2041} {arXiv:0706.2041 [astro-ph]} \BibitemShut
  {NoStop}%
%%CITATION = ARXIV:0706.2041;%%
\bibitem [{\citenamefont {Linder}(2009)}]{Linder:2009jz}%
  \BibitemOpen
  \bibfield  {author} {\bibinfo {author} {\bibfnamefont {E.~V.}\ \bibnamefont
  {Linder}},\ }\href {\doibase 10.1103/PhysRevD.80.123528} {\bibfield
  {journal} {\bibinfo  {journal} {Phys.Rev.}\ }\textbf {\bibinfo {volume}
  {D80}},\ \bibinfo {pages} {123528} (\bibinfo {year} {2009})},\ \Eprint
  {http://arxiv.org/abs/0905.2962} {arXiv:0905.2962 [astro-ph.CO]} \BibitemShut
  {NoStop}%
%%CITATION = ARXIV:0905.2962;%%
\bibitem [{\citenamefont {Cognola}\ \emph {et~al.}(2008)\citenamefont {Cognola}
  \emph {et~al.}}]{Cognola:2007zu}%
  \BibitemOpen
  \bibfield  {author} {\bibinfo {author} {\bibfnamefont {G.}~\bibnamefont
  {Cognola}} \emph {et~al.},\ }\href {\doibase 10.1103/PhysRevD.77.046009}
  {\bibfield  {journal} {\bibinfo  {journal} {Phys.Rev.}\ }\textbf {\bibinfo
  {volume} {D77}},\ \bibinfo {pages} {046009} (\bibinfo {year} {2008})},\
  \Eprint {http://arxiv.org/abs/0712.4017} {arXiv:0712.4017 [hep-th]}
  \BibitemShut {NoStop}%
%%CITATION = ARXIV:0712.4017;%%
\bibitem [{\citenamefont {Deruelle}\ \emph {et~al.}(2008)\citenamefont
  {Deruelle}, \citenamefont {Sasaki},\ and\ \citenamefont
  {Sendouda}}]{Deruelle:2008fs}%
  \BibitemOpen
  \bibfield  {author} {\bibinfo {author} {\bibfnamefont {N.}~\bibnamefont
  {Deruelle}}, \bibinfo {author} {\bibfnamefont {M.}~\bibnamefont {Sasaki}},\
  and\ \bibinfo {author} {\bibfnamefont {Y.}~\bibnamefont {Sendouda}},\ }\href
  {\doibase 10.1103/PhysRevD.77.124024} {\bibfield  {journal} {\bibinfo
  {journal} {Phys.Rev.}\ }\textbf {\bibinfo {volume} {D77}},\ \bibinfo {pages}
  {124024} (\bibinfo {year} {2008})},\ \Eprint {http://arxiv.org/abs/0803.2742}
  {arXiv:0803.2742 [gr-qc]} \BibitemShut {NoStop}%
%%CITATION = ARXIV:0803.2742;%%
\bibitem [{\citenamefont {Tsujikawa}(2008)}]{Tsujikawa:2007xu}%
  \BibitemOpen
  \bibfield  {author} {\bibinfo {author} {\bibfnamefont {S.}~\bibnamefont
  {Tsujikawa}},\ }\href {\doibase 10.1103/PhysRevD.77.023507} {\bibfield
  {journal} {\bibinfo  {journal} {Phys.Rev.}\ }\textbf {\bibinfo {volume}
  {D77}},\ \bibinfo {pages} {023507} (\bibinfo {year} {2008})},\ \Eprint
  {http://arxiv.org/abs/0709.1391} {arXiv:0709.1391 [astro-ph]} \BibitemShut
  {NoStop}%
%%CITATION = ARXIV:0709.1391;%%
\bibitem [{\citenamefont {Hu}\ and\ \citenamefont {Sawicki}(2007)}]{Hu:2007nk}%
  \BibitemOpen
  \bibfield  {author} {\bibinfo {author} {\bibfnamefont {W.}~\bibnamefont
  {Hu}}\ and\ \bibinfo {author} {\bibfnamefont {I.}~\bibnamefont {Sawicki}},\
  }\href {\doibase 10.1103/PhysRevD.76.064004} {\bibfield  {journal} {\bibinfo
  {journal} {Phys.Rev.}\ }\textbf {\bibinfo {volume} {D76}},\ \bibinfo {pages}
  {064004} (\bibinfo {year} {2007})},\ \Eprint {http://arxiv.org/abs/0705.1158}
  {arXiv:0705.1158 [astro-ph]} \BibitemShut {NoStop}%
%%CITATION = ARXIV:0705.1158;%%
\bibitem [{\citenamefont {Li}\ and\ \citenamefont {Barrow}(2007)}]{Li:2007xn}%
  \BibitemOpen
  \bibfield  {author} {\bibinfo {author} {\bibfnamefont {B.}~\bibnamefont
  {Li}}\ and\ \bibinfo {author} {\bibfnamefont {J.~D.}\ \bibnamefont
  {Barrow}},\ }\href {\doibase 10.1103/PhysRevD.75.084010} {\bibfield
  {journal} {\bibinfo  {journal} {Phys.Rev.}\ }\textbf {\bibinfo {volume}
  {D75}},\ \bibinfo {pages} {084010} (\bibinfo {year} {2007})},\ \Eprint
  {http://arxiv.org/abs/gr-qc/0701111} {gr-qc/0701111} \BibitemShut {NoStop}%
%%CITATION = GR-QC/0701111;%%
\bibitem [{\citenamefont {Amendola}\ \emph {et~al.}(2007)\citenamefont
  {Amendola} \emph {et~al.}}]{Amendola:2006we}%
  \BibitemOpen
  \bibfield  {author} {\bibinfo {author} {\bibfnamefont {L.}~\bibnamefont
  {Amendola}} \emph {et~al.},\ }\href {\doibase 10.1103/PhysRevD.75.083504}
  {\bibfield  {journal} {\bibinfo  {journal} {Phys.Rev.}\ }\textbf {\bibinfo
  {volume} {D75}},\ \bibinfo {pages} {083504} (\bibinfo {year} {2007})},\
  \Eprint {http://arxiv.org/abs/gr-qc/0612180} {gr-qc/0612180} \BibitemShut
  {NoStop}%
%%CITATION = GR-QC/0612180;%%
\bibitem [{\citenamefont {Bengochea}\ and\ \citenamefont
  {Ferraro}(2009)}]{Bengochea:2008gz}%
  \BibitemOpen
  \bibfield  {author} {\bibinfo {author} {\bibfnamefont {G.~R.}\ \bibnamefont
  {Bengochea}}\ and\ \bibinfo {author} {\bibfnamefont {R.}~\bibnamefont
  {Ferraro}},\ }\href {\doibase 10.1103/PhysRevD.79.124019} {\bibfield
  {journal} {\bibinfo  {journal} {Phys.Rev.}\ }\textbf {\bibinfo {volume}
  {D79}},\ \bibinfo {pages} {124019} (\bibinfo {year} {2009})},\ \Eprint
  {http://arxiv.org/abs/0812.1205} {arXiv:0812.1205 [astro-ph]} \BibitemShut
  {NoStop}%
%%CITATION = ARXIV:0812.1205;%%
\bibitem [{\citenamefont {Linder}(2010)}]{Linder:2010py}%
  \BibitemOpen
  \bibfield  {author} {\bibinfo {author} {\bibfnamefont {E.~V.}\ \bibnamefont
  {Linder}},\ }\href {\doibase 10.1103/PhysRevD.81.127301} {\bibfield
  {journal} {\bibinfo  {journal} {Phys.Rev.}\ }\textbf {\bibinfo {volume}
  {D81}},\ \bibinfo {pages} {127301} (\bibinfo {year} {2010})},\ \Eprint
  {http://arxiv.org/abs/1005.3039v1} {arXiv:1005.3039v1 [astro-ph.CO]}
  \BibitemShut {NoStop}%
%%CITATION = ARXIV:1005.3039;%%
\bibitem [{\citenamefont {Wei}\ \emph {et~al.}(2011)\citenamefont {Wei},
  \citenamefont {Ma},\ and\ \citenamefont {Qi}}]{Wei:2011jw}%
  \BibitemOpen
  \bibfield  {author} {\bibinfo {author} {\bibfnamefont {H.}~\bibnamefont
  {Wei}}, \bibinfo {author} {\bibfnamefont {X.-P.}\ \bibnamefont {Ma}},\ and\
  \bibinfo {author} {\bibfnamefont {H.-Y.}\ \bibnamefont {Qi}},\ }\href
  {\doibase 10.1016/j.physletb.2011.07.042} {\bibfield  {journal} {\bibinfo
  {journal} {Phys.Lett.}\ }\textbf {\bibinfo {volume} {B703}},\ \bibinfo
  {pages} {74} (\bibinfo {year} {2011})},\ \Eprint
  {http://arxiv.org/abs/1106.0102} {arXiv:1106.0102 [gr-qc]} \BibitemShut
  {NoStop}%
%%CITATION = ARXIV:1106.0102;%%
\bibitem [{\citenamefont {Wei}\ \emph {et~al.}(2012{\natexlab{a}})\citenamefont
  {Wei}, \citenamefont {Qi},\ and\ \citenamefont {Ma}}]{Wei:2011mq}%
  \BibitemOpen
  \bibfield  {author} {\bibinfo {author} {\bibfnamefont {H.}~\bibnamefont
  {Wei}}, \bibinfo {author} {\bibfnamefont {H.-Y.}\ \bibnamefont {Qi}},\ and\
  \bibinfo {author} {\bibfnamefont {X.-P.}\ \bibnamefont {Ma}},\ }\href
  {\doibase 10.1140/epjc/s10052-012-2117-8} {\bibfield  {journal} {\bibinfo
  {journal} {Eur.Phys.J.}\ }\textbf {\bibinfo {volume} {C72}},\ \bibinfo
  {pages} {2117} (\bibinfo {year} {2012}{\natexlab{a}})},\ \Eprint
  {http://arxiv.org/abs/1108.0859} {arXiv:1108.0859 [gr-qc]} \BibitemShut
  {NoStop}%
%%CITATION = ARXIV:1108.0859;%%
\bibitem [{\citenamefont {Wei}\ \emph {et~al.}(2012{\natexlab{b}})\citenamefont
  {Wei}, \citenamefont {Guo},\ and\ \citenamefont {Wang}}]{Wei:2011aa}%
  \BibitemOpen
  \bibfield  {author} {\bibinfo {author} {\bibfnamefont {H.}~\bibnamefont
  {Wei}}, \bibinfo {author} {\bibfnamefont {X.-J.}\ \bibnamefont {Guo}},\ and\
  \bibinfo {author} {\bibfnamefont {L.-F.}\ \bibnamefont {Wang}},\ }\href
  {\doibase 10.1016/j.physletb.2011.12.039} {\bibfield  {journal} {\bibinfo
  {journal} {Phys.Lett.}\ }\textbf {\bibinfo {volume} {B707}},\ \bibinfo
  {pages} {298} (\bibinfo {year} {2012}{\natexlab{b}})},\ \Eprint
  {http://arxiv.org/abs/1112.2270} {arXiv:1112.2270 [gr-qc]} \BibitemShut
  {NoStop}%
%%CITATION = ARXIV:1112.2270;%%
\bibitem [{\citenamefont {Yang}(2011)}]{Yang:2010ji}%
  \BibitemOpen
  \bibfield  {author} {\bibinfo {author} {\bibfnamefont {R.-J.}\ \bibnamefont
  {Yang}},\ }\href {\doibase 10.1209/0295-5075/93/60001} {\bibfield  {journal}
  {\bibinfo  {journal} {Europhys.Lett.}\ }\textbf {\bibinfo {volume} {93}},\
  \bibinfo {pages} {60001} (\bibinfo {year} {2011})},\ \Eprint
  {http://arxiv.org/abs/1010.1376} {arXiv:1010.1376 [gr-qc]} \BibitemShut
  {NoStop}%
%%CITATION = ARXIV:1010.1376;%%
\bibitem [{\citenamefont {Li}\ \emph {et~al.}(2011{\natexlab{a}})\citenamefont
  {Li}, \citenamefont {Sotiriou},\ and\ \citenamefont {Barrow}}]{Li:2010cg}%
  \BibitemOpen
  \bibfield  {author} {\bibinfo {author} {\bibfnamefont {B.}~\bibnamefont
  {Li}}, \bibinfo {author} {\bibfnamefont {T.~P.}\ \bibnamefont {Sotiriou}},\
  and\ \bibinfo {author} {\bibfnamefont {J.~D.}\ \bibnamefont {Barrow}},\
  }\href {\doibase 10.1103/PhysRevD.83.064035} {\bibfield  {journal} {\bibinfo
  {journal} {Phys.Rev.}\ }\textbf {\bibinfo {volume} {D83}},\ \bibinfo {pages}
  {064035} (\bibinfo {year} {2011}{\natexlab{a}})},\ \Eprint
  {http://arxiv.org/abs/1010.1041} {arXiv:1010.1041 [gr-qc]} \BibitemShut
  {NoStop}%
%%CITATION = ARXIV:1010.1041;%%
\bibitem [{\citenamefont {Li}\ \emph {et~al.}(2011{\natexlab{b}})\citenamefont
  {Li}, \citenamefont {Sotiriou},\ and\ \citenamefont {Barrow}}]{Li:2011wu}%
  \BibitemOpen
  \bibfield  {author} {\bibinfo {author} {\bibfnamefont {B.}~\bibnamefont
  {Li}}, \bibinfo {author} {\bibfnamefont {T.~P.}\ \bibnamefont {Sotiriou}},\
  and\ \bibinfo {author} {\bibfnamefont {J.~D.}\ \bibnamefont {Barrow}},\
  }\href {\doibase 10.1103/PhysRevD.83.104017} {\bibfield  {journal} {\bibinfo
  {journal} {Phys.Rev.}\ }\textbf {\bibinfo {volume} {D83}},\ \bibinfo {pages}
  {104017} (\bibinfo {year} {2011}{\natexlab{b}})},\ \Eprint
  {http://arxiv.org/abs/1103.2786} {arXiv:1103.2786 [astro-ph.CO]} \BibitemShut
  {NoStop}%
%%CITATION = ARXIV:1103.2786;%%
\bibitem [{\citenamefont {Wu}\ and\ \citenamefont {Yu}(2011)}]{Wu:2010av}%
  \BibitemOpen
  \bibfield  {author} {\bibinfo {author} {\bibfnamefont {P.-X.}~\bibnamefont
  {Wu}}\ and\ \bibinfo {author} {\bibfnamefont {H.-W.}\ \bibnamefont {Yu}},\
  }\href {\doibase 10.1140/epjc/s10052-011-1552-2} {\bibfield  {journal}
  {\bibinfo  {journal} {Eur.Phys.J.}\ }\textbf {\bibinfo {volume} {C71}},\
  \bibinfo {pages} {1552} (\bibinfo {year} {2011})},\ \Eprint
  {http://arxiv.org/abs/1008.3669} {arXiv:1008.3669 [gr-qc]} \BibitemShut
  {NoStop}%
%%CITATION = ARXIV:1008.3669;%%
\bibitem [{\citenamefont {Nashed}\ and\ \citenamefont
  {El~Hanafy}(2014)}]{Nashed:2014lva}%
  \BibitemOpen
  \bibfield  {author} {\bibinfo {author} {\bibfnamefont {G.}~\bibnamefont
  {Nashed}}\ and\ \bibinfo {author} {\bibfnamefont {W.}~\bibnamefont
  {El~Hanafy}},\ }\href {\doibase 10.1140/epjc/s10052-014-3099-5} {\bibfield
  {journal} {\bibinfo  {journal} {Eur.Phys.J.}\ }\textbf {\bibinfo {volume}
  {C74}},\ \bibinfo {pages} {3099} (\bibinfo {year} {2014})},\ \Eprint
  {http://arxiv.org/abs/1403.0913} {arXiv:1403.0913 [gr-qc]} \BibitemShut
  {NoStop}%
%%CITATION = ARXIV:1403.0913;%%
\bibitem [{\citenamefont {Li}\ \emph {et~al.}(2011{\natexlab{c}})\citenamefont
  {Li}, \citenamefont {Miao},\ and\ \citenamefont {Miao}}]{Li:2011rn}%
  \BibitemOpen
  \bibfield  {author} {\bibinfo {author} {\bibfnamefont {M.}~\bibnamefont
  {Li}}, \bibinfo {author} {\bibfnamefont {R.-X.}\ \bibnamefont {Miao}},\ and\
  \bibinfo {author} {\bibfnamefont {Y.-G.}\ \bibnamefont {Miao}},\ }\href
  {\doibase 10.1007/JHEP07(2011)108} {\bibfield  {journal} {\bibinfo  {journal}
  {JHEP}\ }\textbf {\bibinfo {volume} {1107}},\ \bibinfo {pages} {108}
  (\bibinfo {year} {2011}{\natexlab{c}})},\ \Eprint
  {http://arxiv.org/abs/1105.5934} {arXiv:1105.5934 [hep-th]} \BibitemShut
  {NoStop}%
%%CITATION = ARXIV:1105.5934;%%
\bibitem [{\citenamefont {Nesseris}\ \emph {et~al.}(2013)\citenamefont
  {Nesseris} \emph {et~al.}}]{Nesseris:2013jea}%
  \BibitemOpen
  \bibfield  {author} {\bibinfo {author} {\bibfnamefont {S.}~\bibnamefont
  {Nesseris}} \emph {et~al.},\ }\href {\doibase 10.1103/PhysRevD.88.103010}
  {\bibfield  {journal} {\bibinfo  {journal} {Phys.Rev.}\ }\textbf {\bibinfo
  {volume} {D88}},\ \bibinfo {pages} {103010} (\bibinfo {year} {2013})},\
  \Eprint {http://arxiv.org/abs/1308.6142} {arXiv:1308.6142 [astro-ph.CO]}
  \BibitemShut {NoStop}%
%%CITATION = ARXIV:1308.6142;%%
\bibitem [{\citenamefont
  {Weitzenb{\"o}ck}(1923)}]{weitzenbock1923invariantentheorie}%
  \BibitemOpen
  \bibfield  {author} {\bibinfo {author} {\bibfnamefont {R.}~\bibnamefont
  {Weitzenb{\"o}ck}},\ }\href@noop {} {\emph {\bibinfo {title}
  {{Invariantentheorie}}}}\ (\bibinfo  {publisher} {{Groningen: P.
  Noordhoff}},\ \bibinfo {year} {1923})\BibitemShut {NoStop}%
\bibitem [{\citenamefont {Fierz}\ and\ \citenamefont
  {Pauli}(1939)}]{Fierz:1939ix}%
  \BibitemOpen
  \bibfield  {author} {\bibinfo {author} {\bibfnamefont {M.}~\bibnamefont
  {Fierz}}\ and\ \bibinfo {author} {\bibfnamefont {W.}~\bibnamefont {Pauli}},\
  }\href {\doibase 10.1098/rspa.1939.0140} {\bibfield  {journal} {\bibinfo
  {journal} {Proc.Roy.Soc.Lond.}\ }\textbf {\bibinfo {volume} {A173}},\
  \bibinfo {pages} {211} (\bibinfo {year} {1939})}\BibitemShut {NoStop}%
%%CITATION = PRSLA,A173,211;%%
\bibitem [{\citenamefont {van Dam}\ and\ \citenamefont
  {Veltman}(1970)}]{vanDam:1970vg}%
  \BibitemOpen
  \bibfield  {author} {\bibinfo {author} {\bibfnamefont {H.}~\bibnamefont {van
  Dam}}\ and\ \bibinfo {author} {\bibfnamefont {M.}~\bibnamefont {Veltman}},\
  }\href {\doibase 10.1016/0550-3213(70)90416-5} {\bibfield  {journal}
  {\bibinfo  {journal} {Nucl.Phys.}\ }\textbf {\bibinfo {volume} {B22}},\
  \bibinfo {pages} {397} (\bibinfo {year} {1970})}\BibitemShut {NoStop}%
%%CITATION = NUPHA,B22,397;%%
\bibitem [{\citenamefont {Zakharov}(1970)}]{Zakharov:1970cc}%
  \BibitemOpen
  \bibfield  {author} {\bibinfo {author} {\bibfnamefont {V.}~\bibnamefont
  {Zakharov}},\ }\href@noop {} {\bibfield  {journal} {\bibinfo  {journal} {JETP
  Lett.}\ }\textbf {\bibinfo {volume} {12}},\ \bibinfo {pages} {312} (\bibinfo
  {year} {1970})}\BibitemShut {NoStop}%
%%CITATION = JTPLA,12,312;%%
\bibitem [{\citenamefont {Vainshtein}(1972)}]{Vainshtein:1972sx}%
  \BibitemOpen
  \bibfield  {author} {\bibinfo {author} {\bibfnamefont {A.}~\bibnamefont
  {Vainshtein}},\ }\href {\doibase 10.1016/0370-2693(72)90147-5} {\bibfield
  {journal} {\bibinfo  {journal} {Phys.Lett.}\ }\textbf {\bibinfo {volume}
  {B39}},\ \bibinfo {pages} {393} (\bibinfo {year} {1972})}\BibitemShut
  {NoStop}%
%%CITATION = PHLTA,B39,393;%%
\bibitem [{\citenamefont {Boulware}\ and\ \citenamefont
  {Deser}(1972)}]{Boulware:1973my}%
  \BibitemOpen
  \bibfield  {author} {\bibinfo {author} {\bibfnamefont {D.}~\bibnamefont
  {Boulware}}\ and\ \bibinfo {author} {\bibfnamefont {S.}~\bibnamefont
  {Deser}},\ }\href {\doibase 10.1103/PhysRevD.6.3368} {\bibfield  {journal}
  {\bibinfo  {journal} {Phys.Rev.}\ }\textbf {\bibinfo {volume} {D6}},\
  \bibinfo {pages} {3368} (\bibinfo {year} {1972})}\BibitemShut {NoStop}%
%%CITATION = PHRVA,D6,3368;%%
\bibitem [{\citenamefont {de~Rham}\ and\ \citenamefont
  {Gabadadze}(2010)}]{deRham:2010ik}%
  \BibitemOpen
  \bibfield  {author} {\bibinfo {author} {\bibfnamefont {C.}~\bibnamefont
  {de~Rham}}\ and\ \bibinfo {author} {\bibfnamefont {G.}~\bibnamefont
  {Gabadadze}},\ }\href {\doibase 10.1103/PhysRevD.82.044020} {\bibfield
  {journal} {\bibinfo  {journal} {Phys.Rev.}\ }\textbf {\bibinfo {volume}
  {D82}},\ \bibinfo {pages} {044020} (\bibinfo {year} {2010})},\ \Eprint
  {http://arxiv.org/abs/1007.0443} {arXiv:1007.0443 [hep-th]} \BibitemShut
  {NoStop}%
%%CITATION = ARXIV:1007.0443;%%
\bibitem [{\citenamefont {de~Rham}\ \emph {et~al.}(2011)\citenamefont {de~Rham}
  \emph {et~al.}}]{deRham:2010kj}%
  \BibitemOpen
  \bibfield  {author} {\bibinfo {author} {\bibfnamefont {C.}~\bibnamefont
  {de~Rham}} \emph {et~al.},\ }\href {\doibase 10.1103/PhysRevLett.106.231101}
  {\bibfield  {journal} {\bibinfo  {journal} {Phys.Rev.Lett.}\ }\textbf
  {\bibinfo {volume} {106}},\ \bibinfo {pages} {231101} (\bibinfo {year}
  {2011})},\ \Eprint {http://arxiv.org/abs/1011.1232} {arXiv:1011.1232
  [hep-th]} \BibitemShut {NoStop}%
%%CITATION = ARXIV:1011.1232;%%
\bibitem [{\citenamefont {Hassan}\ and\ \citenamefont
  {Rosen}(2012{\natexlab{a}})}]{Hassan:2011ea}%
  \BibitemOpen
  \bibfield  {author} {\bibinfo {author} {\bibfnamefont {S.}~\bibnamefont
  {Hassan}}\ and\ \bibinfo {author} {\bibfnamefont {R.~A.}\ \bibnamefont
  {Rosen}},\ }\href {\doibase 10.1007/JHEP04(2012)123} {\bibfield  {journal}
  {\bibinfo  {journal} {JHEP}\ }\textbf {\bibinfo {volume} {1204}},\ \bibinfo
  {pages} {123} (\bibinfo {year} {2012}{\natexlab{a}})},\ \Eprint
  {http://arxiv.org/abs/1111.2070} {arXiv:1111.2070 [hep-th]} \BibitemShut
  {NoStop}%
%%CITATION = ARXIV:1111.2070;%%
\bibitem [{\citenamefont {Hassan}\ and\ \citenamefont
  {Rosen}(2012{\natexlab{b}})}]{Hassan:2011hr}%
  \BibitemOpen
  \bibfield  {author} {\bibinfo {author} {\bibfnamefont {S.}~\bibnamefont
  {Hassan}}\ and\ \bibinfo {author} {\bibfnamefont {R.~A.}\ \bibnamefont
  {Rosen}},\ }\href {\doibase 10.1103/PhysRevLett.108.041101} {\bibfield
  {journal} {\bibinfo  {journal} {Phys.Rev.Lett.}\ }\textbf {\bibinfo {volume}
  {108}},\ \bibinfo {pages} {041101} (\bibinfo {year} {2012}{\natexlab{b}})},\
  \Eprint {http://arxiv.org/abs/1106.3344} {arXiv:1106.3344 [hep-th]}
  \BibitemShut {NoStop}%
%%CITATION = ARXIV:1106.3344;%%
\bibitem [{\citenamefont {Motohashi}\ and\ \citenamefont
  {Suyama}(2012)}]{Motohashi:2012jd}%
  \BibitemOpen
  \bibfield  {author} {\bibinfo {author} {\bibfnamefont {H.}~\bibnamefont
  {Motohashi}}\ and\ \bibinfo {author} {\bibfnamefont {T.}~\bibnamefont
  {Suyama}},\ }\href {\doibase 10.1103/PhysRevD.86.081502} {\bibfield
  {journal} {\bibinfo  {journal} {Phys.Rev.}\ }\textbf {\bibinfo {volume}
  {D86}},\ \bibinfo {pages} {081502} (\bibinfo {year} {2012})},\ \Eprint
  {http://arxiv.org/abs/1208.3019} {arXiv:1208.3019 [hep-th]} \BibitemShut
  {NoStop}%
%%CITATION = ARXIV:1208.3019;%%
\bibitem [{\citenamefont {{G{\"u}mr{\"u}k{\c c}{\"u}o{\v g}lu}}\ \emph
  {et~al.}(2012)\citenamefont {{G{\"u}mr{\"u}k{\c c}{\"u}o{\v g}lu}},
  \citenamefont {Lin},\ and\ \citenamefont {Mukohyama}}]{Gumrukcuoglu:2012aa}%
  \BibitemOpen
  \bibfield  {author} {\bibinfo {author} {\bibfnamefont {A.~E.}\ \bibnamefont
  {{G{\"u}mr{\"u}k{\c c}{\"u}o{\v g}lu}}}, \bibinfo {author} {\bibfnamefont
  {C.}~\bibnamefont {Lin}},\ and\ \bibinfo {author} {\bibfnamefont
  {S.}~\bibnamefont {Mukohyama}},\ }\href {\doibase
  10.1016/j.physletb.2012.09.049} {\bibfield  {journal} {\bibinfo  {journal}
  {Phys.Lett.}\ }\textbf {\bibinfo {volume} {B717}},\ \bibinfo {pages} {295}
  (\bibinfo {year} {2012})},\ \Eprint {http://arxiv.org/abs/1206.2723}
  {arXiv:1206.2723 [hep-th]} \BibitemShut {NoStop}%
%%CITATION = ARXIV:1206.2723;%%
\bibitem [{\citenamefont {Kobayashi}\ \emph {et~al.}(2012)\citenamefont
  {Kobayashi} \emph {et~al.}}]{Kobayashi:2012fz}%
  \BibitemOpen
  \bibfield  {author} {\bibinfo {author} {\bibfnamefont {T.}~\bibnamefont
  {Kobayashi}} \emph {et~al.},\ }\href {\doibase 10.1103/PhysRevD.86.061505}
  {\bibfield  {journal} {\bibinfo  {journal} {Phys.Rev.}\ }\textbf {\bibinfo
  {volume} {D86}},\ \bibinfo {pages} {061505} (\bibinfo {year} {2012})},\
  \Eprint {http://arxiv.org/abs/1205.4938} {arXiv:1205.4938 [hep-th]}
  \BibitemShut {NoStop}%
%%CITATION = ARXIV:1205.4938;%%
\bibitem [{\citenamefont {D'Amico}\ \emph {et~al.}(2011)\citenamefont {D'Amico}
  \emph {et~al.}}]{D'Amico:2011jj}%
  \BibitemOpen
  \bibfield  {author} {\bibinfo {author} {\bibfnamefont {G.}~\bibnamefont
  {D'Amico}} \emph {et~al.},\ }\href {\doibase 10.1103/PhysRevD.84.124046}
  {\bibfield  {journal} {\bibinfo  {journal} {Phys.Rev.}\ }\textbf {\bibinfo
  {volume} {D84}},\ \bibinfo {pages} {124046} (\bibinfo {year} {2011})},\
  \Eprint {http://arxiv.org/abs/1108.5231} {arXiv:1108.5231 [hep-th]}
  \BibitemShut {NoStop}%
%%CITATION = ARXIV:1108.5231;%%
\bibitem [{\citenamefont {De~Felice}\ \emph {et~al.}(2012)\citenamefont
  {De~Felice} \emph {et~al.}}]{DeFelice:2012mx}%
  \BibitemOpen
  \bibfield  {author} {\bibinfo {author} {\bibfnamefont {A.}~\bibnamefont
  {De~Felice}} \emph {et~al.},\ }\href {\doibase
  10.1103/PhysRevLett.109.171101} {\bibfield  {journal} {\bibinfo  {journal}
  {Phys.Rev.Lett.}\ }\textbf {\bibinfo {volume} {109}},\ \bibinfo {pages}
  {171101} (\bibinfo {year} {2012})},\ \Eprint {http://arxiv.org/abs/1206.2080}
  {arXiv:1206.2080 [hep-th]} \BibitemShut {NoStop}%
%%CITATION = ARXIV:1206.2080;%%
\bibitem [{\citenamefont {De~Felice}\ \emph
  {et~al.}(2013{\natexlab{a}})\citenamefont {De~Felice} \emph
  {et~al.}}]{DeFelice:2013bxa}%
  \BibitemOpen
  \bibfield  {author} {\bibinfo {author} {\bibfnamefont {A.}~\bibnamefont
  {De~Felice}} \emph {et~al.},\ }\href {\doibase
  10.1088/0264-9381/30/18/184004} {\bibfield  {journal} {\bibinfo  {journal}
  {Class.Quant.Grav.}\ }\textbf {\bibinfo {volume} {30}},\ \bibinfo {pages}
  {184004} (\bibinfo {year} {2013}{\natexlab{a}})},\ \Eprint
  {http://arxiv.org/abs/1304.0484} {arXiv:1304.0484 [hep-th]} \BibitemShut
  {NoStop}%
%%CITATION = ARXIV:1304.0484;%%
\bibitem [{\citenamefont {De~Felice}\ \emph
  {et~al.}(2013{\natexlab{b}})\citenamefont {De~Felice} \emph
  {et~al.}}]{DeFelice:2013awa}%
  \BibitemOpen
  \bibfield  {author} {\bibinfo {author} {\bibfnamefont {A.}~\bibnamefont
  {De~Felice}} \emph {et~al.},\ }\href {\doibase 10.1088/1475-7516/2013/05/035}
  {\bibfield  {journal} {\bibinfo  {journal} {JCAP}\ }\textbf {\bibinfo
  {volume} {1305}},\ \bibinfo {pages} {035} (\bibinfo {year}
  {2013}{\natexlab{b}})},\ \Eprint {http://arxiv.org/abs/1303.4154}
  {arXiv:1303.4154 [hep-th]} \BibitemShut {NoStop}%
%%CITATION = ARXIV:1303.4154;%%
\bibitem [{\citenamefont {Huang}\ \emph {et~al.}(2012)\citenamefont {Huang},
  \citenamefont {Piao},\ and\ \citenamefont {Zhou}}]{Huang:2012pe}%
  \BibitemOpen
  \bibfield  {author} {\bibinfo {author} {\bibfnamefont {Q.-G.}\ \bibnamefont
  {Huang}}, \bibinfo {author} {\bibfnamefont {Y.-S.}\ \bibnamefont {Piao}},\
  and\ \bibinfo {author} {\bibfnamefont {S.-Y.}\ \bibnamefont {Zhou}},\ }\href
  {\doibase 10.1103/PhysRevD.86.124014} {\bibfield  {journal} {\bibinfo
  {journal} {Phys.Rev.}\ }\textbf {\bibinfo {volume} {D86}},\ \bibinfo {pages}
  {124014} (\bibinfo {year} {2012})},\ \Eprint {http://arxiv.org/abs/1206.5678}
  {arXiv:1206.5678 [hep-th]} \BibitemShut {NoStop}%
%%CITATION = ARXIV:1206.5678;%%
\bibitem [{\citenamefont {Wu}\ \emph {et~al.}(2013)\citenamefont {Wu},
  \citenamefont {Piao},\ and\ \citenamefont {Cai}}]{Wu:2013ii}%
  \BibitemOpen
  \bibfield  {author} {\bibinfo {author} {\bibfnamefont {D.-J.}\ \bibnamefont
  {Wu}}, \bibinfo {author} {\bibfnamefont {Y.-S.}\ \bibnamefont {Piao}},\ and\
  \bibinfo {author} {\bibfnamefont {Y.-F.}\ \bibnamefont {Cai}},\ }\href
  {\doibase 10.1016/j.physletb.2013.02.055} {\bibfield  {journal} {\bibinfo
  {journal} {Phys.Lett.}\ }\textbf {\bibinfo {volume} {B721}},\ \bibinfo
  {pages} {7} (\bibinfo {year} {2013})},\ \Eprint
  {http://arxiv.org/abs/1301.4326} {arXiv:1301.4326 [hep-th]} \BibitemShut
  {NoStop}%
%%CITATION = ARXIV:1301.4326;%%
\bibitem [{\citenamefont {D'Amico}\ \emph
  {et~al.}(2013{\natexlab{a}})\citenamefont {D'Amico} \emph
  {et~al.}}]{D'Amico:2012zv}%
  \BibitemOpen
  \bibfield  {author} {\bibinfo {author} {\bibfnamefont {G.}~\bibnamefont
  {D'Amico}} \emph {et~al.},\ }\href {\doibase 10.1103/PhysRevD.87.064037}
  {\bibfield  {journal} {\bibinfo  {journal} {Phys.Rev.}\ }\textbf {\bibinfo
  {volume} {D87}},\ \bibinfo {pages} {064037} (\bibinfo {year}
  {2013}{\natexlab{a}})},\ \Eprint {http://arxiv.org/abs/1206.4253}
  {arXiv:1206.4253 [hep-th]} \BibitemShut {NoStop}%
%%CITATION = ARXIV:1206.4253;%%
\bibitem [{\citenamefont {D'Amico}\ \emph
  {et~al.}(2013{\natexlab{b}})\citenamefont {D'Amico} \emph
  {et~al.}}]{D'Amico:2013kya}%
  \BibitemOpen
  \bibfield  {author} {\bibinfo {author} {\bibfnamefont {G.}~\bibnamefont
  {D'Amico}} \emph {et~al.},\ }\href {\doibase 10.1088/0264-9381/30/18/184005}
  {\bibfield  {journal} {\bibinfo  {journal} {Class.Quant.Grav.}\ }\textbf
  {\bibinfo {volume} {30}},\ \bibinfo {pages} {184005} (\bibinfo {year}
  {2013}{\natexlab{b}})},\ \Eprint {http://arxiv.org/abs/1304.0723}
  {arXiv:1304.0723 [hep-th]} \BibitemShut {NoStop}%
%%CITATION = ARXIV:1304.0723;%%
\bibitem [{\citenamefont {De~Felice}\ \emph
  {et~al.}(2013{\natexlab{c}})\citenamefont {De~Felice} \emph
  {et~al.}}]{DeFelice:2013dua}%
  \BibitemOpen
  \bibfield  {author} {\bibinfo {author} {\bibfnamefont {A.}~\bibnamefont
  {De~Felice}} \emph {et~al.},\ }\href {\doibase 10.1103/PhysRevD.88.124006}
  {\bibfield  {journal} {\bibinfo  {journal} {Phys.Rev.}\ }\textbf {\bibinfo
  {volume} {D88}},\ \bibinfo {pages} {124006} (\bibinfo {year}
  {2013}{\natexlab{c}})},\ \Eprint {http://arxiv.org/abs/1309.3162}
  {arXiv:1309.3162 [hep-th]} \BibitemShut {NoStop}%
%%CITATION = ARXIV:1309.3162;%%
\bibitem [{\citenamefont {Cai}\ \emph {et~al.}(2014)\citenamefont {Cai},
  \citenamefont {Duplessis},\ and\ \citenamefont {Saridakis}}]{Cai:2013lqa}%
  \BibitemOpen
  \bibfield  {author} {\bibinfo {author} {\bibfnamefont {Y.-F.}\ \bibnamefont
  {Cai}}, \bibinfo {author} {\bibfnamefont {F.}~\bibnamefont {Duplessis}},\
  and\ \bibinfo {author} {\bibfnamefont {E.~N.}\ \bibnamefont {Saridakis}},\
  }\href {\doibase 10.1103/PhysRevD.90.064051} {\bibfield  {journal} {\bibinfo
  {journal} {Phys.Rev.}\ }\textbf {\bibinfo {volume} {D90}},\ \bibinfo {pages}
  {064051} (\bibinfo {year} {2014})},\ \Eprint {http://arxiv.org/abs/1307.7150}
  {arXiv:1307.7150 [hep-th]} \BibitemShut {NoStop}%
%%CITATION = ARXIV:1307.7150;%%
\bibitem [{\citenamefont {Cai}\ and\ \citenamefont
  {Saridakis}(2014)}]{Cai:2014upa}%
  \BibitemOpen
  \bibfield  {author} {\bibinfo {author} {\bibfnamefont {Y.-F.}\ \bibnamefont
  {Cai}}\ and\ \bibinfo {author} {\bibfnamefont {E.~N.}\ \bibnamefont
  {Saridakis}},\ }\href {\doibase 10.1103/PhysRevD.90.063528} {\bibfield
  {journal} {\bibinfo  {journal} {Phys.Rev.}\ }\textbf {\bibinfo {volume}
  {D90}},\ \bibinfo {pages} {063528} (\bibinfo {year} {2014})},\ \Eprint
  {http://arxiv.org/abs/1401.4418} {arXiv:1401.4418 [astro-ph.CO]} \BibitemShut
  {NoStop}%
%%CITATION = ARXIV:1401.4418;%%
\bibitem [{\citenamefont {Einstein}(1928{\natexlab{a}})}]{einstein1928riemann}%
  \BibitemOpen
  \bibfield  {author} {\bibinfo {author} {\bibfnamefont {A.}~\bibnamefont
  {Einstein}},\ }\href@noop {} {\bibfield  {journal} {\bibinfo  {journal}
  {{Sitzber. Preuss. Akad. Wiss.}}\ }\textbf {\bibinfo {volume} {{17}}},\
  \bibinfo {pages} {217} (\bibinfo {year} {1928}{\natexlab{a}})}\BibitemShut
  {NoStop}%
\bibitem [{\citenamefont {Einstein}(1928{\natexlab{b}})}]{einstein1928neue}%
  \BibitemOpen
  \bibfield  {author} {\bibinfo {author} {\bibfnamefont {A.}~\bibnamefont
  {Einstein}},\ }\href@noop {} {\bibfield  {journal} {\bibinfo  {journal}
  {{Sitzber. Preuss. Akad. Wiss.}}\ }\textbf {\bibinfo {volume} {{17}}},\
  \bibinfo {pages} {224} (\bibinfo {year} {1928}{\natexlab{b}})}\BibitemShut
  {NoStop}%
\bibitem [{\citenamefont {Einstein}(1930)}]{einstein1930auf}%
  \BibitemOpen
  \bibfield  {author} {\bibinfo {author} {\bibfnamefont {A.}~\bibnamefont
  {Einstein}},\ }\href {\doibase 10.1007/BF01782370} {\bibfield  {journal}
  {\bibinfo  {journal} {{Math. Annal.}}\ }\textbf {\bibinfo {volume} {{102}}},\
  \bibinfo {pages} {685} (\bibinfo {year} {1930})}\BibitemShut {NoStop}%
\bibitem [{\citenamefont {Unzicker}\ and\ \citenamefont
  {Case}(2005)}]{Unzicker:2005in}%
  \BibitemOpen
  \bibfield  {author} {\bibinfo {author} {\bibfnamefont {A.}~\bibnamefont
  {Unzicker}}\ and\ \bibinfo {author} {\bibfnamefont {T.}~\bibnamefont
  {Case}},}\href@noop {} \ \Eprint
  {http://arxiv.org/abs/physics/0503046} {physics/0503046} \BibitemShut
  {NoStop}%
%%CITATION = PHYSICS/0503046;%%
\bibitem [{\citenamefont {Hayashi}\ and\ \citenamefont
  {Shirafuji}(1979)}]{Hayashi:1979qx}%
  \BibitemOpen
  \bibfield  {author} {\bibinfo {author} {\bibfnamefont {K.}~\bibnamefont
  {Hayashi}}\ and\ \bibinfo {author} {\bibfnamefont {T.}~\bibnamefont
  {Shirafuji}},\ }\href {\doibase 10.1103/PhysRevD.19.3524} {\bibfield
  {journal} {\bibinfo  {journal} {Phys.Rev.}\ }\textbf {\bibinfo {volume}
  {D19}},\ \bibinfo {pages} {3524} (\bibinfo {year} {1979})}\BibitemShut
  {NoStop}%
%%CITATION = PHRVA,D19,3524;%%
\bibitem [{\citenamefont {Ferraro}\ and\ \citenamefont
  {Fiorini}(2007)}]{Ferraro:2006jd}%
  \BibitemOpen
  \bibfield  {author} {\bibinfo {author} {\bibfnamefont {R.}~\bibnamefont
  {Ferraro}}\ and\ \bibinfo {author} {\bibfnamefont {F.}~\bibnamefont
  {Fiorini}},\ }\href {\doibase 10.1103/PhysRevD.75.084031} {\bibfield
  {journal} {\bibinfo  {journal} {Phys.Rev.}\ }\textbf {\bibinfo {volume}
  {D75}},\ \bibinfo {pages} {084031} (\bibinfo {year} {2007})},\ \Eprint
  {http://arxiv.org/abs/gr-qc/0610067} {gr-qc/0610067} \BibitemShut {NoStop}%
%%CITATION = GR-QC/0610067;%%
\bibitem [{\citenamefont {Ferraro}\ and\ \citenamefont
  {Fiorini}(2008)}]{Ferraro:2008ey}%
  \BibitemOpen
  \bibfield  {author} {\bibinfo {author} {\bibfnamefont {R.}~\bibnamefont
  {Ferraro}}\ and\ \bibinfo {author} {\bibfnamefont {F.}~\bibnamefont
  {Fiorini}},\ }\href {\doibase 10.1103/PhysRevD.78.124019} {\bibfield
  {journal} {\bibinfo  {journal} {Phys.Rev.}\ }\textbf {\bibinfo {volume}
  {D78}},\ \bibinfo {pages} {124019} (\bibinfo {year} {2008})},\ \Eprint
  {http://arxiv.org/abs/0812.1981} {arXiv:0812.1981 [gr-qc]} \BibitemShut
  {NoStop}%
%%CITATION = ARXIV:0812.1981;%%
\bibitem [{\citenamefont {Hinterbichler}\ and\ \citenamefont
  {Rosen}(2012)}]{Hinterbichler:2012cn}%
  \BibitemOpen
  \bibfield  {author} {\bibinfo {author} {\bibfnamefont {K.}~\bibnamefont
  {Hinterbichler}}\ and\ \bibinfo {author} {\bibfnamefont {R.~A.}\ \bibnamefont
  {Rosen}},\ }\href {\doibase 10.1007/JHEP07(2012)047} {\bibfield  {journal}
  {\bibinfo  {journal} {JHEP}\ }\textbf {\bibinfo {volume} {1207}},\ \bibinfo
  {pages} {047} (\bibinfo {year} {2012})},\ \Eprint
  {http://arxiv.org/abs/1203.5783} {arXiv:1203.5783 [hep-th]} \BibitemShut
  {NoStop}%
%%CITATION = ARXIV:1203.5783;%%
\bibitem [{\citenamefont {Suzuki}\ \emph {et~al.}(2012)\citenamefont {Suzuki}
  \emph {et~al.}}]{Suzuki:2011hu}%
  \BibitemOpen
  \bibfield  {author} {\bibinfo {author} {\bibfnamefont {N.}~\bibnamefont
  {Suzuki}} \emph {et~al.},\ }\href {\doibase 10.1088/0004-637X/746/1/85}
  {\bibfield  {journal} {\bibinfo  {journal} {Astrophys.J.}\ }\textbf {\bibinfo
  {volume} {746}},\ \bibinfo {pages} {85} (\bibinfo {year} {2012})},\ \Eprint
  {http://arxiv.org/abs/1105.3470} {arXiv:1105.3470 [astro-ph.CO]} \BibitemShut
  {NoStop}%
%%CITATION = ARXIV:1105.3470;%%
\bibitem [{\citenamefont {Nesseris}\ and\ \citenamefont
  {Perivolaropoulos}(2005)}]{Nesseris:2005ur}%
  \BibitemOpen
  \bibfield  {author} {\bibinfo {author} {\bibfnamefont {S.}~\bibnamefont
  {Nesseris}}\ and\ \bibinfo {author} {\bibfnamefont {L.}~\bibnamefont
  {Perivolaropoulos}},\ }\href {\doibase 10.1103/PhysRevD.72.123519} {\bibfield
   {journal} {\bibinfo  {journal} {Phys.Rev.}\ }\textbf {\bibinfo {volume}
  {D72}},\ \bibinfo {pages} {123519} (\bibinfo {year} {2005})},\ \Eprint
  {http://arxiv.org/abs/astro-ph/0511040} {arXiv:astro-ph/0511040} \BibitemShut
  {NoStop}%
%%CITATION = ASTRO-PH/0511040;%%
\bibitem [{\citenamefont {Wang}\ and\ \citenamefont
  {Mukherjee}(2006)}]{Wang:2006ts}%
  \BibitemOpen
  \bibfield  {author} {\bibinfo {author} {\bibfnamefont {Y.}~\bibnamefont
  {Wang}}\ and\ \bibinfo {author} {\bibfnamefont {P.}~\bibnamefont
  {Mukherjee}},\ }\href {\doibase 10.1086/507091} {\bibfield  {journal}
  {\bibinfo  {journal} {Astrophys.J.}\ }\textbf {\bibinfo {volume} {650}},\
  \bibinfo {pages} {1} (\bibinfo {year} {2006})},\ \Eprint
  {http://arxiv.org/abs/astro-ph/0604051} {arXiv:astro-ph/0604051} \BibitemShut
  {NoStop}%
%%CITATION = ASTRO-PH/0604051;%%
\bibitem [{\citenamefont {Bond}\ \emph {et~al.}(1997)\citenamefont {Bond},
  \citenamefont {Efstathiou},\ and\ \citenamefont {Tegmark}}]{Bond:1997wr}%
  \BibitemOpen
  \bibfield  {author} {\bibinfo {author} {\bibfnamefont {J.}~\bibnamefont
  {Bond}}, \bibinfo {author} {\bibfnamefont {G.}~\bibnamefont {Efstathiou}},\
  and\ \bibinfo {author} {\bibfnamefont {M.}~\bibnamefont {Tegmark}},\
  }\href@noop {} {\bibfield  {journal} {\bibinfo  {journal}
  {Mon.Not.Roy.Astron.Soc.}\ }\textbf {\bibinfo {volume} {291}},\ \bibinfo
  {pages} {L33} (\bibinfo {year} {1997})},\ \Eprint
  {http://arxiv.org/abs/astro-ph/9702100} {astro-ph/9702100} \BibitemShut
  {NoStop}%
%%CITATION = ASTRO-PH/9702100;%%
\bibitem [{\citenamefont {Ade}\ \emph {et~al.}(2015{\natexlab{a}})\citenamefont
  {Ade} \emph {et~al.}}]{Planck:2015xua}%
  \BibitemOpen
  \bibfield  {author} {\bibinfo {author} {\bibfnamefont {P.}~\bibnamefont
  {Ade}} \emph {et~al.} (\bibinfo {collaboration} {Planck Collaboration}),\
  }\href@noop {} \Eprint
  {http://arxiv.org/abs/1502.01589} {arXiv:1502.01589 [astro-ph.CO]}
  \BibitemShut {NoStop}%
%%CITATION = ARXIV:1502.01589;%%
\bibitem [{\citenamefont {Ade}\ \emph {et~al.}(2015{\natexlab{b}})\citenamefont
  {Ade} \emph {et~al.}}]{Ade:2015rim}%
  \BibitemOpen
  \bibfield  {author} {\bibinfo {author} {\bibfnamefont {P.}~\bibnamefont
  {Ade}} \emph {et~al.} (\bibinfo {collaboration} {Planck Collaboration}),\
  }\href@noop {} \Eprint
  {http://arxiv.org/abs/1502.01590} {arXiv:1502.01590 [astro-ph.CO]}
  \BibitemShut {NoStop}%
%%CITATION = ARXIV:1502.01590;%%
\bibitem [{\citenamefont {Eisenstein}\ \emph {et~al.}(2005)\citenamefont
  {Eisenstein} \emph {et~al.}}]{Eisenstein:2005su}%
  \BibitemOpen
  \bibfield  {author} {\bibinfo {author} {\bibfnamefont {D.~J.}\ \bibnamefont
  {Eisenstein}} \emph {et~al.} (\bibinfo {collaboration} {SDSS
  Collaboration}),\ }\href {\doibase 10.1086/466512} {\bibfield  {journal}
  {\bibinfo  {journal} {Astrophys.J.}\ }\textbf {\bibinfo {volume} {633}},\
  \bibinfo {pages} {560} (\bibinfo {year} {2005})},\ \Eprint
  {http://arxiv.org/abs/astro-ph/0501171} {astro-ph/0501171} \BibitemShut
  {NoStop}%
%%CITATION = ASTRO-PH/0501171;%%
\bibitem [{\citenamefont {Jassal}\ \emph {et~al.}(2010)\citenamefont {Jassal}
  \emph {et~al.}}]{Jassal:2006gf}%
  \BibitemOpen
  \bibfield  {author} {\bibinfo {author} {\bibfnamefont {H.~K.}\ \bibnamefont
  {Jassal}} \emph {et~al.},\ }\href {\doibase 10.1111/j.1365-2966.2010.16647.x}
  {\bibfield  {journal} {\bibinfo  {journal} {Mon.Not.Roy.Astron.Soc.}\
  }\textbf {\bibinfo {volume} {405}},\ \bibinfo {pages} {2639} (\bibinfo {year}
  {2010})},\ \Eprint {http://arxiv.org/abs/astro-ph/0601389} {astro-ph/0601389}
  \BibitemShut {NoStop}%
%%CITATION = ASTRO-PH/0601389;%%
\bibitem [{\citenamefont {Alam}\ \emph {et~al.}(2004)\citenamefont {Alam},
  \citenamefont {Sahni},\ and\ \citenamefont {Starobinsky}}]{Alam:2004jy}%
  \BibitemOpen
  \bibfield  {author} {\bibinfo {author} {\bibfnamefont {U.}~\bibnamefont
  {Alam}}, \bibinfo {author} {\bibfnamefont {V.}~\bibnamefont {Sahni}},\ and\
  \bibinfo {author} {\bibfnamefont {A.}~\bibnamefont {Starobinsky}},\ }\href
  {\doibase 10.1088/1475-7516/2004/06/008} {\bibfield  {journal} {\bibinfo
  {journal} {JCAP}\ }\textbf {\bibinfo {volume} {0406}},\ \bibinfo {pages}
  {008} (\bibinfo {year} {2004})},\ \Eprint
  {http://arxiv.org/abs/astro-ph/0403687} {astro-ph/0403687} \BibitemShut
  {NoStop}%
%%CITATION = ASTRO-PH/0403687;%%
\bibitem [{\citenamefont {Nesseris}\ and\ \citenamefont
  {Perivolaropoulos}(2007)}]{Nesseris:2006er}%
  \BibitemOpen
  \bibfield  {author} {\bibinfo {author} {\bibfnamefont {S.}~\bibnamefont
  {Nesseris}}\ and\ \bibinfo {author} {\bibfnamefont {L.}~\bibnamefont
  {Perivolaropoulos}},\ }\href {\doibase 10.1088/1475-7516/2007/01/018}
  {\bibfield  {journal} {\bibinfo  {journal} {JCAP}\ }\textbf {\bibinfo
  {volume} {0701}},\ \bibinfo {pages} {018} (\bibinfo {year} {2007})},\ \Eprint
  {http://arxiv.org/abs/astro-ph/0610092} {astro-ph/0610092} \BibitemShut
  {NoStop}%
%%CITATION = ASTRO-PH/0610092;%%
\bibitem [{\citenamefont {Wu}\ and\ \citenamefont {Yu}(2006)}]{Wu:2006bb}%
  \BibitemOpen
  \bibfield  {author} {\bibinfo {author} {\bibfnamefont {P.-X.}\ \bibnamefont
  {Wu}}\ and\ \bibinfo {author} {\bibfnamefont {H.-W.}\ \bibnamefont {Yu}},\
  }\href {\doibase 10.1016/j.physletb.2006.11.021} {\bibfield  {journal}
  {\bibinfo  {journal} {Phys.Lett.}\ }\textbf {\bibinfo {volume} {B643}},\
  \bibinfo {pages} {315} (\bibinfo {year} {2006})},\ \Eprint
  {http://arxiv.org/abs/astro-ph/0611507} {astro-ph/0611507} \BibitemShut
  {NoStop}%
%%CITATION = ASTRO-PH/0611507;%%
\bibitem [{\citenamefont {Bamba}\ \emph {et~al.}(2011)\citenamefont {Bamba}
  \emph {et~al.}}]{Bamba:2010wb}%
  \BibitemOpen
  \bibfield  {author} {\bibinfo {author} {\bibfnamefont {K.}~\bibnamefont
  {Bamba}} \emph {et~al.},\ }\href {\doibase 10.1088/1475-7516/2011/01/021}
  {\bibfield  {journal} {\bibinfo  {journal} {JCAP}\ }\textbf {\bibinfo
  {volume} {1101}},\ \bibinfo {pages} {021} (\bibinfo {year} {2011})},\ \Eprint
  {http://arxiv.org/abs/1011.0508} {arXiv:1011.0508 [astro-ph.CO]} \BibitemShut
  {NoStop}%
%%CITATION = ARXIV:1011.0508;%%
\bibitem [{\citenamefont {Wu}\ and\ \citenamefont {Yu}(2010)}]{Wu:2010mn}%
  \BibitemOpen
  \bibfield  {author} {\bibinfo {author} {\bibfnamefont {P.-X.}~\bibnamefont
  {Wu}}\ and\ \bibinfo {author} {\bibfnamefont {H.-W.}\ \bibnamefont {Yu}},\
  }\href {\doibase 10.1016/j.physletb.2010.08.073} {\bibfield  {journal}
  {\bibinfo  {journal} {Phys.Lett.}\ }\textbf {\bibinfo {volume} {B693}},\
  \bibinfo {pages} {415} (\bibinfo {year} {2010})},\ \Eprint
  {http://arxiv.org/abs/1006.0674} {arXiv:1006.0674 [gr-qc]} \BibitemShut
  {NoStop}%
%%CITATION = ARXIV:1006.0674;%%
\bibitem [{\citenamefont {Bamba}\ \emph {et~al.}(2010)\citenamefont {Bamba},
  \citenamefont {Geng},\ and\ \citenamefont {Lee}}]{Bamba:2010iw}%
  \BibitemOpen
  \bibfield  {author} {\bibinfo {author} {\bibfnamefont {K.}~\bibnamefont
  {Bamba}}, \bibinfo {author} {\bibfnamefont {C.-Q.}\ \bibnamefont {Geng}},\
  and\ \bibinfo {author} {\bibfnamefont {C.-C.}\ \bibnamefont {Lee}},\
  }\href@noop {} \Eprint
  {http://arxiv.org/abs/1008.4036} {arXiv:1008.4036 [astro-ph.CO]} \BibitemShut
  {NoStop}%
%%CITATION = ARXIV:1008.4036;%%
%%%%%%%%%%%%%%%%%%%%%%%%%%%%%%%%%%%%%%%%%%%%%%%%%%%%%% v2: Refs. added by HW
%%CITATION = ARXIV:hep-th/0601213;%%
\bibitem [{\citenamefont {Nojiri}\ and\ \citenamefont {Odintsov}(2007)}]
  {Nojiri:2006ri}%
  \BibitemOpen
  \bibfield  {author} {\bibinfo {author} {\bibfnamefont {S.}~\bibnamefont
  {Nojiri}}\ and\ \bibinfo {author} {\bibfnamefont {S.~D.}\ \bibnamefont
  {Odintsov}},\ }\href {\doibase 10.1142/S0219887807001928}
  {\bibfield  {journal}
  {\bibinfo  {journal} {Int.J.Geom.Meth.Mod.Phys.}\ }\textbf {\bibinfo
  {volume} {4}},\ \bibinfo {pages} {115} (\bibinfo {year} {2007})},
  \Eprint {http://arxiv.org/abs/hep-th/0601213} {hep-th/0601213} \BibitemShut
  {NoStop}%
%%CITATION = ARXIV:hep-th/0601213;%%
%%CITATION = ARXIV:1011.0544;%%
\bibitem [{\citenamefont {Nojiri}\ and\ \citenamefont {Odintsov}(2011)}]
  {Nojiri:2010wj}%
  \BibitemOpen
  \bibfield  {author} {\bibinfo {author} {\bibfnamefont {S.}~\bibnamefont
  {Nojiri}}\ and\ \bibinfo {author} {\bibfnamefont {S.~D.}\ \bibnamefont
  {Odintsov}},\ }\href {\doibase 10.1016/j.physrep.2011.04.001}
  {\bibfield  {journal}
  {\bibinfo  {journal} {Phys.Rept.}\ }\textbf {\bibinfo
  {volume} {505}},\ \bibinfo {pages} {115} (\bibinfo {year} {2011})},
  \Eprint {http://arxiv.org/abs/arXiv:1011.0544} {arXiv:1011.0544 [gr-qc]}
  \BibitemShut {NoStop}%
%%CITATION = ARXIV:1011.0544;%%
%%CITATION = ARXIV:0710.1738;%%
\bibitem [{\citenamefont {Nojiri}\ and\ \citenamefont {Odintsov}(2008)}]
  {Nojiri:2007cq}%
  \BibitemOpen
  \bibfield  {author} {\bibinfo {author} {\bibfnamefont {S.}~\bibnamefont
  {Nojiri}}\ and\ \bibinfo {author} {\bibfnamefont {S.~D.}\ \bibnamefont
  {Odintsov}},\ }\href {\doibase 10.1103/PhysRevD.77.026007}
  {\bibfield  {journal}
  {\bibinfo  {journal} {Phys.Rev.}\ }\textbf {\bibinfo
  {volume} {D77}},\ \bibinfo {pages} {026007} (\bibinfo {year} {2008})},
  \Eprint {http://arxiv.org/abs/arXiv:0710.1738} {arXiv:0710.1738 [hep-th]}
  \BibitemShut {NoStop}%
%%CITATION = ARXIV:0710.1738;%%
%%CITATION = ARXIV:0707.1941;%%
\bibitem [{\citenamefont {Nojiri}\ and\ \citenamefont {Odintsov}(2007)}]
  {Nojiri:2007as}%
  \BibitemOpen
  \bibfield  {author} {\bibinfo {author} {\bibfnamefont {S.}~\bibnamefont
  {Nojiri}}\ and\ \bibinfo {author} {\bibfnamefont {S.~D.}\ \bibnamefont
  {Odintsov}},\ }\href {\doibase 10.1016/j.physletb.2007.10.027}
  {\bibfield  {journal}
  {\bibinfo  {journal} {Phys.Lett.}\ }\textbf {\bibinfo
  {volume} {B657}},\ \bibinfo {pages} {238} (\bibinfo {year} {2007})},
  \Eprint {http://arxiv.org/abs/arXiv:0707.1941} {arXiv:0707.1941 [hep-th]}
  \BibitemShut {NoStop}%
%%CITATION = ARXIV:0707.1941;%%
%%CITATION = ARXIV:1309.2185;%%
\bibitem [{\citenamefont {Kluson}\ \emph {et~al.}(2013)\citenamefont {Kluson}
  \emph {et~al.}}]{Kluson:2013yaa}%
  \BibitemOpen
  \bibfield  {author} {\bibinfo {author} {\bibfnamefont {J.}~\bibnamefont
  {Kluso\v{n}}},\ \bibinfo {author} {\bibfnamefont {S.}~\bibnamefont
  {Nojiri}},\ and\ \bibinfo {author} {\bibfnamefont {S.~D.}\ \bibnamefont
  {Odintsov}},\ }\href {\doibase 10.1016/j.physletb.2013.10.003}
  {\bibfield  {journal}
  {\bibinfo  {journal} {Phys.Lett.}\ }\textbf {\bibinfo
  {volume} {B726}},\ \bibinfo {pages} {918} (\bibinfo {year} {2013})},
  \Eprint {http://arxiv.org/abs/arXiv:1309.2185} {arXiv:1309.2185 [hep-th]}
  \BibitemShut {NoStop}%
%%CITATION = ARXIV:1309.2185;%%
\end{thebibliography}%

\end{document}